\documentclass[%
 reprint,
nofootinbib,
 amsmath,amssymb,
 aps,
]{revtex4-2}

\usepackage{dcolumn}% Align table columns on decimal point
\usepackage{bm}% bold math
\usepackage{amsmath}
\usepackage[dvipdfmx]{graphicx}
\usepackage{epstopdf}
\usepackage{epsfig}
\usepackage{amssymb}
\usepackage{braket}
\usepackage{comment}
\usepackage{color}
\usepackage{multirow} %表を縦に結合する

\begin{document}

\preprint{APS/123-QED}

\title{Stability of the embedded string in the $SU(N)\times U(1)$ Higgs model \\ and its application}% Force line breaks with \\

\author{Yukihiro Kanda}
 %\altaffiliation[Also at ]{Department of Physics,
 %Nagoya University, Nagoya 464-8602, Japan}%Lines break automatically or can be forced with \\
 \email{kanda.y@eken.phys.nagoya-u.ac.jp}
\affiliation{%
Department of Physics,
Nagoya University, Nagoya 464-8602, Japan
}%


\author{Nobuhiro Maekawa}
\email{maekawa@eken.phys.nagoya-u.ac.jp}
 \affiliation{%
 Department of Physics,
 Nagoya University, Nagoya 464-8602, Japan
 }%
\affiliation{
  Kobayashi-Maskawa Institute for the Origin of Particles and the
  Universe, Nagoya University, Nagoya 464-8602, Japan
}%
% \affiliation{%
%  Authors' institution and/or address\\
%  This line break forced with \textbackslash\textbackslash
% }%

% \collaboration{CLEO Collaboration}%\noaffiliation

\date{\today}% It is always \today, today,

\begin{abstract}
  Since it has been pointed out that physics beyond the Standard Model may be constrained by gravitational waves from cosmic strings, it has been more important to clarify in what cases cosmic strings are formed. 
  We study the stability of the embedded string which is formed when $SU(N)\times U(1)_X$ gauge symmetries are broken to $SU(N-1)\times U(1)_Q$, and find that the stability condition
can be determined by two mass ratios of the Higgs and massive gauge bosons, and
does not explicitly depend on $N$. 
We also show that the result can be extended in supersymmetric models. 
In addition, we apply these results to several models and discuss the important feature of the Higgs to produce the embedded string. Although we find it difficult to be satisfied
in normal realistic GUT models, it is possible if $SU(N)$ and $U(1)_X$ have different origins.
% \begin{description}
% Secondary publications and information retrieval purposes.
% You may use the \texttt{description} environment to structure your abstract;
% use the optional argument of the \verb+\item+ command to give the category of each item. 
% \end{description}
\end{abstract}

\maketitle

%\tableofcontents

\section{Introduction}

It is known that cosmic strings\cite{Kibble:1976sj,Vilenkin:2000jqa} are formed as topological defects after phase transitions in  a lot of models beyond the Standard Model (SM) 
including grand unified theories (GUTs)\cite{Kibble:1982ae,Jeannerot:2003qv,Dvali:1993sg} . 
The characteristic signatures of these strings can be observed through   
cosmic microwave background\cite{Albrecht:1997nt}, gravitational lensing\cite{Vilenkin:1984ea} or gravitational wave background\cite{Damour:2000wa}.
Moreover, the observation of the gravitational wave spectrum reveals the tension of the cosmic string, which gives the energy scale of the phase transition
\cite{Blanco-Pillado:2017oxo,Auclair:2019wcv}. In 2020, NANOGrav experiment reported their result
\cite{NANOGrav:2020bcs}, which is consistent with gravitational wave signal from cosmic strings\cite{Ellis:2020ena,Blasi:2020mfx,Buchmuller:2020lbh}. This signal suggests the presence of a symmetry breaking whose energy scale is $10^{14-16}$ GeV\cite{Blasi:2020mfx,Chigusa:2020rks}. Furthermore, several gravitational waves observations, for example LISA\cite{Amaro-Seoane:2012aqc} and DECIGO\cite{Seto:2001qf}, are planned within a few decades. Since the gravitational wave observations will be powerful tools for detecting past phase transitions\cite{Auclair:2019wcv}, 
it is important to clarify the conditions for cosmic
string formation.

There exists a well known result for cosmic string formation.
When $U(1)$ gauge symmetry is broken by developing a vacuum expectation value (VEV) of a complex Higgs, a cosmic string, which is called Nielsen-Olesen string (N-O string), can be formed\cite{Nielsen:1973cs}. The existence of the N-O string is related to a topological feature of a moduli space $\mathcal{V}$, specifically the first homotopy group $\pi_1(\mathcal{V})$. This N-O string can be generalized in other spontaneous symmetry breaking (SSB). If $\pi_1(\mathcal{V})$ is nontrivial for the general SSB, stable cosmic strings can be produced\cite{Kibble:1976sj}. Since the stability of these strings are guaranteed by the topological features of the moduli space, these strings are called  topological strings.

The above general argument for cosmic string formation does not mean that cosmic strings cannot be formed when $\pi_1(\mathcal{V})$ is trivial.
Actually, the string solutions in the electroweak symmetry
breaking, in which $\pi_1(\mathcal{V})$ is trivial, have been studied.
These are called the Z-strings 
(for a review, see Ref.\cite{Achucarro:1999it}). 
The idea of the Z-string solution with the ends (the electroweak dumbbell) has been proposed Ref.\cite{Nambu:1977ag} by Nambu, and the Z-string solution without the ends has been considered in Ref.\cite{Vachaspati:1992fi} by Vachaspati. The Z-string is not always classically stable and its stability depends on two parameters, the mixing angle $\theta_W$ and the ratio of Higgs mass to Z boson mass $m_H/m_Z$. The region in the space of the two parameters in which the Z-string is classically stable has been calculated numerically in Ref.\cite{James:1992zp}, and it has been clear that the Z-string becomes unstable with the realistic parameters in the SM. The stability of the Z-string has been studied in the two Higgs doublet model (2HDM) and has been found to be also unstable in the realistic parameters in the SM\cite{La:1993je,Earnshaw:1993yu,Perivolaropoulos:1993gg}. The Z-string solution is constructed by embedding the N-O string in the Higgs doublet field and the Z gauge boson field. The strings constructed 
by embedding the N-O strings are called embedded strings\cite{Vachaspati:1992pi} and
the Z-string is one of them.  

As mentioned above, the embedded strings have been well studied in the symmetry breaking, $SU(2)_L\times U(1)_Y\rightarrow U(1)_Q$, but it is not yet clear whether they are formed or not in other SSB. There are many predicted models beyond the SM, such as GUTs, and they have various SSB at high energy scale. 
It becomes more important to clarify which SSB produces embedded strings or not, because the gravitational waves from the embedded strings may be detected in future experiments.

In this paper, we consider more general embedded string which may be produced in the gauge symmetry breaking $SU(N)\times U(1)_X\rightarrow SU(N-1)\times U(1)_Q$ and examine its classical stability. 
Since this is a generalization of the Z-string, we call the embedded string as generalized Z-string in this paper. The method to check the stability is the same as  in Ref.\cite{James:1992zp}, thus we calculate the sign of the energy variation made by infinitesimal perturbations. As a result, we find that the classical stability of the generalized Z-string is essentially determined by two mass ratios of Higgs and massive gauge bosons.
To achieve the stability, the mass of neutral gauge boson must be at least several times the mass of the charged gauge boson.
We also consider the generalized Z-string in the supersymmetric (SUSY) $SU(N)\times U(1)_X$ Higgs model and show that its classical stability is essentially the same as in the non-SUSY case. For $N=2$, it is pointed out in Ref.\cite{Kanda:2022xrz}.

Since the breaking $SU(N)\times U(1)_X\rightarrow SU(N-1)\times U(1)_Q$ can be seen in many GUT models, we apply the condition for generalized string formation to several scenarios of GUTs in which $SU(N)\times U(1)_X$ breaking happens.
Unfortunately, we conclude that it is difficult to satisfy the condition in normal realistic GUT models. However, if a special GUT which we explain later can be constructed, the condition may be satisfied.
To obtain several times larger neutral gauge boson mass than the charged gauge boson mass, the Higgs must have large $U(1)_X$ charge which become possible if the Higgs belong to higher representation field of the unified gauge group. 
We discuss how large a representation field, in which the Higgs field is included, we need to produce classically stable generalized Z-string in several toy GUT models.
If the embedded strings are discovered by future cosmological observations, it may become a strict constraint for various models beyond the SM.

This paper is organized as follows. In Sec.2, we review the Z-string and how to check its classical stability because we use the same method for our study. We examine the classical stability of the generalized Z-string in the $SU(N)\times U(1)_X$ model in Sec.3 and consider its SUSY extension in Sec.4. As a result of them, we find the condition for a formation of the generalized Z-string. In Sec.5, we apply  the condition to the case in which $SU(N)$ and $U(1)_X$ are unified into a simple group.

%%%%%%%%%%%%%%%%%%%%%%%%%%%%%%%%
\section{REVIEW OF THE Z-STRING}
%%%%%%%%%%%%%%%%%%%%%%%%%%%%%%%%

In this section, we will review the Z-string and how to check its stability briefly. The Z-string is an embedded string which can be constructed in a gauge theory with $SU(2)_L\times U(1)_Y$ gauge symmetry broken to $U(1)_Q$ by developing a VEV of a doublet Higgs. 

First, let us show the concrete form of the Z-string solution which has been found by Vachaspati\cite{Vachaspati:1992fi}. We consider $SU(2)_L\times U(1)_Y$ gauge theory with a doublet Higgs $H$ which has $U(1)_Y$ charge $1/2$. The Lagrangian is given as
\begin{align}
    \label{L_EW}
    \mathcal{L} = &-\frac{1}{4}W^a_{\mu\nu}W^{a\mu\nu} - \frac{1}{4}B_{\mu\nu}B^{\mu\nu} \nonumber\\
    &+ \left| D_\mu H \right|^2 - \lambda \left(\left|H\right|^2 - v^2\right)^2,
\end{align}
where $W^a_{\mu\nu}$ $(a=1,2,3)$ and $B_{\mu\nu}$ are field strengths of $SU(2)_L$ and $U(1)_Y$, respectively. In this model, the gauge symmetries are broken to $U(1)_Q$ when $H$ obtains a non-vanishing VEV. 
When we take $W_\mu^a$ and $B_\mu$ as the gauge fields of $SU(2)_L$ and $U(1)_Y$, respectively, 
the gauge field for unbroken $U(1)_Q$ is $A_\mu \equiv \sin\theta_W W^3_\mu + \cos\theta_W B_\mu$ and those for the broken gauge symmetries are
\begin{align}
    W^1_\mu, \quad W^2_\mu, \quad Z_\mu \equiv \cos\theta_W W^3_\mu - \sin\theta_W B_\mu,
\end{align} 
where $g_1$ and $g_2$ denote the gauge coupling constants of $U(1)_Y$ and $SU(2)_L$, respectively, and $\tan \theta_W \equiv g_1/g_2$. The moduli space of Higgs for this breaking is homeomorphic to $S^3$, and hence there is no topological string formed ($\because\, \pi_1(S^3)$ is trivial). However, an embedded string can be formed as shown in the following. 

The Z-string solutions are classical solutions of this system and they are given as
\begin{align}
    \label{Z-string solution}
    \begin{aligned}
        &H(x) = \left(\begin{array}{c}
            0 \\ f(r) e^{in\theta}
        \end{array}\right), \quad Z_\theta(x) = - n z(r), \\
        &Z_t(x) = Z_r(x) = Z_z(x) = A_\mu(x) = W^{\bar{a}}_\mu(x) = 0 \\
        &(\bar{a} = 1, 2),
    \end{aligned}
\end{align}
where we use cylindrical coordinates $(t, r, \theta, z)$ and $n\in \mathbb{Z}\setminus \{0\}$ is a winding number. $f(r)$ and $z(r)$ are monotonic increasing functions of $r$ which satisfy boundary conditions
\begin{align}
    f(0) = z(0) = 0, \quad f(\infty) = v, \quad z(\infty) = \frac{2}{\alpha} ,
\end{align}
where $\alpha\equiv \sqrt{g_1^2+g_2^2}$. The shapes of them are determined by the Euler-Lagrange equations which are obtained from the Lagrangian (\ref{L_EW}) %which are given 
as
\begin{align}
    \label{feq_EW}
    &f''(r) + \frac{f'(r)}{r} -n^2\left(1 - \frac{\alpha}{2}z(r)\right)^2 \frac{f(r)}{r^2} \nonumber\\
    &\qquad\qquad\qquad+ 2\lambda\left(v^2 - f(r)^2\right) f(r) = 0 , \\
    \label{zeq_EW}
    &z''(r) - \frac{z'(r)}{r} + \alpha\left(1-\frac{\alpha}{2}z(r)\right)f^2(r) = 0 .
\end{align}
These equations can be solved numerically.

The first homotopy group of the moduli space of $H$ is trivial, thus any topological strings such as N-O strings do not appear in the breaking, $SU(2)_L\times U(1)_Y \rightarrow U(1)_Q$. In the notation of $Z_\mu, A_\mu$ and $W_\mu^{\bar{a}}$, we can deform the covariant derivative as
\begin{align}
    &D_\mu H = \left(\partial_\mu + i \frac{\alpha}{2} Z_\mu T_Z -i\frac{g_2g_1}{\alpha}A_\mu T_A - ig_2 W^{\bar{a}}_\mu \frac{\sigma^{\bar{a}}}{2}\right) H \nonumber\\
    &(\bar{a} = 1, 2), 
\end{align}
where
\begin{align}
    T_Z \equiv \left(\begin{array}{cc}
        \sin^2\theta_W - \cos^2\theta_W & 0 \\ 0 & 1
    \end{array}\right), \quad
    T_A \equiv \left(\begin{array}{cc}
        1&0\\0&0
    \end{array}\right) .
\end{align}
Now we show $U(1)$ which is generated by $T_z$ as $U(1)_Z$. If we ignore $W_\mu^{\bar{a}}$ and $A_\mu$, we can regard the symmetry breaking as $U(1)_Z\rightarrow \times$ and find "N-O string" solutions which is related to this $U(1)_Z$ breaking. They are nothing but the Z-string solutions. Thus, the Z-string is constructed as N-O string for symmetry breaking of the subgroup, and called an embedded string.

The Z-string is a classical solution, but it is not sure that it is classically stable. Next, we show how to check the classical stability of the Z-string. The method is very simple. We check whether perturbation modes of the Z-string solutions make the energy of the system lower or not. The perturbation modes of the $n=1$ Z-string solutions are given as
\begin{align}
    \begin{aligned}
        &H(x) = \left(\begin{array}{c}
            h(x) \\ f(r) e^{i\theta} + \delta\phi(x)
        \end{array}\right), \quad Z_0(x) = \delta Z_0(x), \\
        &\vec{Z}(x) = - \frac{z(r)}{r}\vec{e}_\theta + \delta\vec{Z}(x)
    \end{aligned}
\end{align}
and we also consider $A_\mu(x)$ and $W^{\bar{a}}_\mu(x)$ as perturbations. We substitute them into the energy of the system and evaluate the sign of the variation. If the variation becomes negative, there is a perturbation mode which makes the Z-string solutions unstable. 

Since the $t$ and $z$ dependence of the perturbations and non-vanishing $t$ and $z$ components of the gauge fields only increases the energy, we take the perturbations independent of $t$ and $z$, and we ignore the $t$ and $z$ components of the gauge fields. Thus, we can determine the stability using an energy linear density (string tension) $\mu_{EW}$ instead of the energy. $\mu_{EW}$ is given as
\begin{align}
    \label{mu_EW}
    \mu_{EW} = \int rdrd\theta &\left[ \frac{1}{4} \left( W^a_{\bar{i}\bar{j}} \right)^2 + \frac{1}{4} \left( B_{\bar{i}\bar{j}} \right)^2 + \left| D_{\bar{i}} H \right|^2 \right.\nonumber\\
    &\left.+ \lambda \left(\left|H\right|^2 - v^2\right)^2 \right] \qquad (\bar{i}, \bar{j} = 1, 2).
\end{align}

Next, we find perturbations which do not give negative variation and ignore them. 
Since the Z-string solution satisfies the classical equations of motion, the leading terms of the variation of the string tension are quadratic terms of these
perturbation modes $\delta\phi(x)$, $\delta Z_\mu(x)$, $h(x)$, $A_\mu(x)$ and $W^{\bar{a}}_\mu(x)$. Because of conservation of $U(1)_Q$ charge, the quadratic terms of neutral fields 
$\delta\mu_n$ and the quadratic terms of charged fields $\delta\mu_c$ are separated, i.e., $\delta\mu=\mu-\mu_0\sim \delta\mu_n+\delta\mu_c$. The neutral part $\delta\mu_n$ must be non-negative because this part takes the same form as the perturbation from the energy linear density of the N-O string solution in the $U(1)$ Higgs model. This has been also checked numerically in Ref.\cite{Goodband:1995he}. Thus, it is sufficient to evaluate the sign of $\delta\mu_c$, which is given as
\begin{widetext}
\begin{align}
    \label{mu_c}
    \delta\mu_c &= \int d^2x \left[ 2\left\{ \left(\nabla\times\vec{W}^{+}\right)_z -\frac{ig_2^2}{\alpha}\frac{z}{r}W^{+}_r \right\} \left\{ \left(\nabla\times\vec{W}^{-}\right)_z + \frac{ig_2^2}{\alpha}\frac{z}{r}W^{-}_r \right\}
    + \frac{2ig_2^2}{\alpha} \frac{z'}{r} \left(\vec{W}^+ \times \vec{W}^- \right)_z  \right.\nonumber\\
    &\hspace{17mm} -ig_2 \left\{ \left(f' W^{-}_r -if \frac{(1-\frac{\alpha}{2}z)}{r} W^{-}_\theta\right)e^{-i\theta} h + f e^{i\theta} \left(W^{+}_r \partial_rh^* + W^{+}_\theta \frac{1}{r}\left(\partial_\theta - i\frac{\alpha}{2}\left(\cos^2\theta_W - \sin^2\theta_W\right)z\right)h^*\right) \right\} \nonumber\\
    &\hspace{17mm} +ig_2 \left\{ \left(f' W^{+}_r +if \frac{(1-\frac{\alpha}{2}z)}{r} W^{+}_\theta\right)e^{i\theta} h^* + f e^{-i\theta} \left(W^{-}_r \partial_r h + W^{-}_\theta \frac{1}{r}\left(\partial_\theta + i\frac{\alpha}{2}\left(\cos^2\theta_W - \sin^2\theta_W\right)z\right)h\right)  \right\} \nonumber\\
    &\hspace{17mm}\left. + \left|\partial_r h \right|^2 + \frac{1}{r^2}\left| \left(\partial_\theta + i\frac{\alpha}{2} \left(\cos^2\theta_W - \sin^2\theta_W\right) z\right) h \right|^2 + g_2^2f^2 W^{+}_{\bar{i}}W^{-}_{\bar{i}} + 2\lambda\left(f^2 -v^2\right)|h|^2 \right] ,
\end{align}
\end{widetext}
where
\begin{align}
    \left(\begin{array}{c}
        \vec{W}^{+} \\ \vec{W}^{-}
    \end{array}\right)
    \equiv \frac{1}{2}
    \left(\begin{array}{ccc}
        1 && -i \\ 1 && i
    \end{array}\right)
    \left(\begin{array}{c}
        \vec{W}^1 \\ \vec{W}^2
    \end{array}\right) .
\end{align}

Ignoring other perturbation modes which do not make the Z-string solutions unstable, it can be seen that there is only one perturbation mode that could give a negative contribution. The perturbation mode can be written as a certain linear combination of $h(x)$ and $W_{\bar{i}}^{\bar{a}}(x)$. The calculations to reduce the number of perturbation modes are summarized in the appendix.
\footnote{In the appendix, we calculate for the case of $SU(N)\times U(1)\rightarrow SU(N-1)\times U(1)$. This is just a generalization of the electroweak breaking and set N=2 if you want to see the Z-string case.} 
The variation of string tension is given as
\begin{align}
    \label{mu_zeta_EW}
    \delta\mu_\zeta &= 2\pi \int rdr \, \zeta \left[ -\frac{1}{r} \frac{d}{dr}\left(\frac{r}{P}\frac{d}{dr}\right) \right. \nonumber\\
    &\hspace{20mm}\left.+ \left\{ \frac{2S}{r^2f^2} + \frac{1}{r}\frac{d}{dr}\left(\frac{rf'}{Pf}\right) + \frac{(f')^2}{Pf^2} \right\} \right]\zeta \nonumber\\
    &\equiv 2\pi\int rdr \, \zeta \mathcal{O} \zeta ,
\end{align}
where
\begin{align}
    P = \frac{g_2^2r^2f^2}{2} + \left(1- \frac{g_2^2}{\alpha} z\right)^2,
\end{align}
\begin{align}
    S = r\frac{d}{dr}\left(\frac{1}{rP}\left(1-\frac{g_2^2}{\alpha}z\right)\frac{z'}{\alpha}\right) + \frac{f^2}{2} -\frac{1}{P}\frac{g_2^2}{\alpha^2}(z')^2 .
\end{align}
In Eq.(\ref{mu_zeta_EW}), we have already integrated by $\theta$ and $\zeta(r)$ corresponds to the perturbation mode. If the operator $\mathcal{O}$ has a negative eigenvalue, there must be a perturbation which makes $\delta\mu_\zeta$ negative. Therefore, it is important to know whether $\mathcal{O}$ has negative eigenvalues or not. It has already been solved by numerical calculation and we also review it briefly for later discussions. To do the numerical calculation, $r, f(r)$ and $z(r)$ are non-dimensionalized as
\begin{align}
    R \equiv \frac{\alpha v}{2}r, \quad F(R) \equiv \frac{f(r)}{v} ,\quad Z(R) \equiv \frac{\alpha}{2} z(r) .
\end{align}
In this non-dimensionalization, the Euler-Lagrange equations, Eqs.(\ref{feq_EW}) and (\ref{zeq_EW}), are rewritten as
\begin{align}
    \label{EoM_F_EW}
    &F''(R) + \frac{F'(R)}{R} - \left(1-Z(R)\right)^2\frac{F(R)}{R^2} \nonumber\\
    &\hspace{25mm}+ \beta\left(1-F^2(R)\right)F(R) =0 ,\\
    \label{EoM_Z_EW}
    &Z''(R) - \frac{Z'(R)}{R} + 2(1-Z(R))F^2(R) = 0 .
\end{align}
where $\beta\equiv 8\lambda/\alpha^2$. $\mathcal{O}$ is also normalized as
\begin{align}
    \label{o_tilde}
    \mathcal{O} &= \frac{\alpha v^2}{4} \left[ -\frac{1}{R} \frac{d}{dR}\left(\frac{R}{\tilde{P}}\frac{d}{dR}\right) \right. \nonumber\\
    &\hspace{12mm}\left.+ \left\{ \frac{2\tilde{S}}{R^2F^2} + \frac{1}{R}\frac{d}{dR}\left(\frac{RF'}{\tilde{P}F}\right) + \frac{(F')^2}{\tilde{P}F^2} \right\} \right] \nonumber\\
    &\equiv \frac{\alpha v^2}{4} \tilde{\mathcal{O}} ,
\end{align}
where
\begin{align}
    \tilde{P} &\equiv P = 2 \cos^2\theta_W R^2 F^2 + \left(1 - 2\cos^2\theta_W Z\right)^2 \\
    \tilde{S} &\equiv \frac{S}{v^2} = \frac{R}{2}\frac{d}{dR} \left( \frac{Z'}{R\tilde{P}}\left(1-2\cos^2\theta_W Z\right) \right) \nonumber\\
    &\hspace{20mm}+ \frac{F^2}{2} - \frac{1}{\tilde{P}}\cos^2\theta_W (Z')^2 .
\end{align}
Since $\tilde{\mathcal{O}}$ only depends on $\beta$ and $\cos^2\theta_W$, the condition for the Z-string solutions to be stable are given as a region in parameter space $(\beta, \cos^2\theta_W)$, which is shown in Fig. 1 in Ref.\cite{James:1992zp}.
Basically, $\cos^2\theta_W\sim 0$ is required. This is consistent with the stability of 
semi-local string string in the model with $SU(2)$ global symmetry $\times U(1)$ gauge symmetry (i.e., $g_2=0$ in Z-string)\cite{Vachaspati:1991dz}.

%%%%%%%%%%%%%%%%%%%%%%%%%%%%%%%%%%%
\section{Embedded string in $SU(N)\times U(1)_X\rightarrow SU(N-1)\times U(1)_Q$}
%%%%%%%%%%%%%%%%%%%%%%%%%%%%%%%%%%%

Embedded string solutions can also exist in other breaking than the electroweak symmetry breaking\cite{Vachaspati:1992pi}. In this section, we consider the embedded string in a breaking where $SU(N)\times U(1)_X$ gauge symmetries are broken to $SU(N-1)\times U(1)_Q$. It is a generalization of the Z-string.

First, we consider the $SU(N)\times U(1)_X$ gauge theory with $SU(N)$ fundamental Higgs $\phi$ whose $U(1)_Q$ charge is normalized as $1/2$.
The Lagrangian is given as
\begin{align}
    \mathcal{L} = -\frac{1}{4}G^a_{\mu\nu}G^{a\mu\nu} - \frac{1}{4}F_{\mu\nu}F^{\mu\nu} + \left| D_\mu \phi \right|^2 - \lambda \left(\left|\phi\right|^2 - v^2\right)^2 ,
\end{align}
where $G^a_{\mu\nu}\, (a=1,\dots ,N^2-1)$ and $F_{\mu\nu}$ are field strengths of $SU(N)$ and $U(1)_X$, respectively. The covariant derivative of $\phi$ is
\begin{align}
    D_\mu \phi = \left(\partial_\mu - i g_N G^a_\mu T^a_N - i \frac{g_1}{2} F_\mu\right) \phi
\end{align}
where $G^a_\mu$ and $F_\mu$ are gauge fields of $SU(N)$ and $U(1)_X$, respectively. $T^a_N$ are generator matrices of the fundamental representation of $SU(N)$ which satisfy that $\mbox{tr}\left[T^aT^b\right]=\delta^{ab}/2$. In this paper, we choose the basis of fundamental representation such that $T^a_N$ are given as follows.
\begin{align}
    \left(T_N^{\alpha}\right)_{ij} &= \left\{ \begin{aligned}
        \left( T_{N-1}^{\alpha}\right)_{ij} \qquad (i,j\leq N-1) \\
        0 \qquad (i=N \,\mbox{or}\, j=N)
    \end{aligned}\right. \\
    & \qquad (\alpha=1,\dots ,(N-1)^2-1) \nonumber\\
    \left(T_N^{\bar{a}}\right)_{ij} &= \frac{1}{2}\delta_{i,\frac{\bar{a}-(N-1)^2}{2}+1}\delta_{j,N} + \frac{1}{2}\delta_{i,N}\delta_{j,\frac{\bar{a}-(N-1)^2}{2}+1}\delta_{j,N} 
     \nonumber\\ &\hspace{3mm} - \frac{i}{2}\delta_{i,\frac{\bar{a}-(N-1)^2+1}{2}}\delta_{j,N} + \frac{i}{2}\delta_{i,N}\delta_{j,\frac{\bar{a}-(N-1)^2+1}{2}}\delta_{j,N} \nonumber\\
     &\qquad (\bar{a}=(N-1)^2,\dots ,N^2-2) \\
     T_N^{N^2-1} &= \frac{1}{\sqrt{2N(N-1)}}\times\mbox{diag}(1,\dots , 1, 1-N) ,
\end{align}
Approximately, these are also given as 
\begin{widetext}
\begin{align}
    &T^{\alpha}_N = \left(\begin{array}{ccc|c}
         &&&\vdots  \\ &T_{N-1}^{\alpha}&&0 \\ &&&\vdots \\ \hline \cdots&0&\cdots&0
    \end{array}\right) , \quad
    T^{N^2-1}_N = \frac{1}{\sqrt{2N(N-1)}}\left(\begin{array}{ccc|c}
        1&&&\vdots  \\ &\ddots&&0 \\ &&1&\vdots \\ \hline \cdots&0&\cdots&1-N
    \end{array}\right) \nonumber \\
    &T^{\bar{a}}_N = \frac{1}{2}\left(\begin{array}{ccccccc|c}
         &&&&&&& 0 \\ &&&&&&& \vdots \\ &&&&&&&0 \\ &&&\mbox{\huge{0}}&&&&1 \\ &&&&&&&0 \\ &&&&&&&\vdots \\ &&&&&&& 0  \\\hline 0&\cdots&0& 1 &0&\cdots&0&0
    \end{array}\right) \quad\mbox{or}\quad
    \frac{1}{2}\left(\begin{array}{ccccccc|c}
        &&&&&&& 0 \\ &&&&&&& \vdots \\ &&&&&&&0 \\ &&&\mbox{\huge{0}}&&&&-i \\ &&&&&&&0 \\ &&&&&&&\vdots \\ &&&&&&& 0  \\\hline 0&\cdots&0& i &0&\cdots&0&0
   \end{array}\right) .
\end{align}
\end{widetext}

When $\phi$ obtains VEV as $\braket{\phi}= (0,\dots ,0,v)^\top$, $SU(N)\times U(1)_X$ are broken to $SU(N-1)\times U(1)_Q$. The massless gauge fields are 
\begin{align}
    G^{\alpha}_\mu,\quad \tilde{A}_\mu \equiv \frac{g_1^2}{c_N^2g_N^2 + g_1^2} G^{N^2-1}_\mu + \frac{c_N^2g_N^2}{c_N^2g_N^2 + g_1^2} F_\mu ,
\end{align}
and the massive gauge fields are
\begin{align}
    G^{\bar{a}}_\mu,\quad \tilde{Z}_\mu \equiv \frac{c_N^2g_N^2}{c_N^2g_N^2 + g_1^2} G^{N^2-1}_\mu - \frac{g_1^2}{c_N^2g_N^2 + g_1^2} F_\mu ,
\end{align}
where $c_N = \sqrt{\frac{2(N-1)}{N}}$. To simplify notation, we define $\alpha_N \equiv \sqrt{g_1^2 + c_N^2 g_N^2}$ and $\tan\theta_G \equiv g_1/(c_N g_N)$. Thus the relation between $(G^{N^2-1}_\mu, F_\mu)$ and $(\tilde{Z}_\mu, \tilde{A}_\mu)$ are given as
\begin{align}
    \left( \begin{array}{c}
        \tilde{Z}_\mu \\ \tilde{A}_\mu
    \end{array}\right)
    =
    \left(\begin{array}{cc}
        \cos\theta_G & -\sin\theta_G \\ \sin\theta_G & \cos\theta_G
    \end{array}\right)
    \left( \begin{array}{c}
        G^{N^2-1}_\mu \\ F_\mu
    \end{array}\right) .
\end{align}
Particulary when $N=2$, this model is nothing but the $SU(2)_L\times U(1)_Y$ Higgs model.

For a later discussion, we rewrite the coupling constants using the ratio of the masses. After the SSB, there are one Higgs, one neutral gauge boson and $2(N-1)$ charged gauge bosons as massive modes. Their squared masses are given as
\begin{align}
    &m_\phi^2 = 8\lambda v^2 ,\quad m_{\tilde{Z}}^2 = \alpha_N^2 v^2 = (g_1^2+c_N^2g_N^2) v^2 , \nonumber\\
    &m_G^2 = g_N^2 v^2 ,
\end{align}
respectively. Thus we can rewrite the coupling constants as 
\begin{align}
    \label{coupling_mass}
    g_N \rightarrow \frac{m_G}{v} ,\quad g_1 \rightarrow \frac{\sqrt{m_{\tilde{Z}}^2 - c_N^2 m_G^2}}{v} ,\quad \lambda \rightarrow \frac{m_\phi^2}{8v^2} .
\end{align}
Hereafter, we write the coupling constants as in (\ref{coupling_mass}). This reparameterization will makes the discussion about the stability of the embedded string clear as we will show later.

The moduli space of Higgs is homeomorphic to $S^{2N-1}$ and no topological string is formed. However, if we consider N-O string solutions in the breaking of $U(1)$ which corresponds to the gauge transformation of $\tilde{Z}_\mu$, they are nothing but an embedded string solution in $SU(N)\times U(1)_X\rightarrow SU(N-1)\times U(1)_Q$. Because $\tilde{Z}_\mu$ does not have charge of the $U(1)_Q$, the solutions are generalized solutions of Z-string solutions. Hence we call them ``generalized Z-string'' solutions in this paper. The ansatz of them is given as
\begin{align}
    \begin{aligned}
        \label{gZstring_ansatz}
        &\phi(x) = \left(\begin{array}{c}
            0 \\ \vdots \\ 0 \\ f(r) e^{in\theta}
        \end{array}\right) , \quad
        \tilde{Z}_\theta(x) = - nz(r) , \quad (n\in \mathbb{Z}\setminus \{0\})\\
        &\tilde{Z}_t(x) = \tilde{Z}_r(x) = \tilde{Z}_z(x) = \tilde{A}_\mu(x) = G^{\alpha}_\mu = G^{\bar{a}}_\mu =0 .
    \end{aligned} 
\end{align}
For this solution to be solitonic, they have to make the energy density nonzero when $r\sim0$ and zero when $r\rightarrow\infty$. The energy density of the generalized Z-string is given as
\begin{align}
    \label{e_density}
    \mathcal{E} = \frac{n^2 {z'}^2}{2r^2} + {f'}^2 + \frac{n^2f^2}{r^2} \left(1-\frac{m_{\tilde{Z}}}{2v}z\right)^2 + \frac{m_\phi^2}{8v^2} \left(v^2-f^2\right)^2 .
\end{align}
Thus, $f(r)$ and $z(r)$ satisfy the boundary conditions,
\begin{align}
    f(0) = z(0) = 0, \quad f(\infty)=v, \quad z(\infty)=\frac{2v}{m_{\tilde{Z}}} .
\end{align}
Because Eq.(\ref{gZstring_ansatz}) are classical solutions of this system, we can find equations which $f$ and $z$ obey by substituting Eq.(\ref{gZstring_ansatz}) into the Euler-Lagrange equation. The equations are given as
\begin{align}
    \label{feq_SUN}
    &f''(r) + \frac{f'(r)}{r} -n^2\left(1 - \frac{m_{\tilde{Z}}}{2v}z(r)\right)^2 \frac{f(r)}{r^2} \nonumber\\
    &\hspace{30mm}+ \frac{m_\phi^2}{4v^2}\left(v^2 - f(r)^2\right) f(r) = 0 , \\
    \label{zeq_SUN}
    &z''(r) - \frac{z'(r)}{r} + \frac{m_{\tilde{Z}}}{v}\left(1-\frac{m_{\tilde{Z}}}{2v}z(r)\right)f^2(r) = 0 .
\end{align}
Furthermore, if we non-dimensionalize  $r, f(r)$ and $z(r)$ as
\begin{align}
    \label{normalize_SUN}
    R \equiv \frac{m_{\tilde{Z}}}{2}r ,\quad F(R) \equiv \frac{f(r)}{v} ,\quad Z(R) \equiv \frac{m_{\tilde{Z}}}{2v} z(r) ,
\end{align}
the boundary conditions become
\begin{align}
    \label{Boundary}
    F(0) = Z(0) = 0, \quad F(\infty)=Z(\infty)=1,
\end{align}
and Eqs.(\ref{feq_SUN}) and (\ref{zeq_SUN}) are rewritten as
\begin{align}
    \label{EoM_F_SUN}
    &F''(R) + \frac{F'(R)}{R} - \left(1-Z(R)\right)^2\frac{F(R)}{R^2} \nonumber\\
    &\hspace{30mm}+ \frac{m_\phi^2}{m_{\tilde{Z}}^2}\left(1-F^2(R)\right)F(R) =0 \\
    \label{EoM_Z_SUN}
    &Z''(R) - \frac{Z'(R)}{R} + 2(1-Z(R))F^2(R) = 0 .
\end{align}
Note that these non-dimensionalized equations and boundary conditions 
do not explicitly depend on $N$. 
Hence we find that the shape of $F(R)$ and $Z(R)$ depend only on the ratio of masses of Higgs and the massive neutral gauge boson.

Next, we explore conditions that the generalized Z-string solutions are classically stable as
in Sec. 2. 
The perturbation modes around the generalized Z-string are denoted as
\begin{align}
    \delta\phi(x), \quad \delta\tilde{Z}_\mu(x), \quad \phi_c(x), \quad G_\mu^{\alpha}(x), \quad G_\mu^{\bar{a}}(x), \quad \tilde{A}_\mu(x) ,
\end{align}
where
\begin{align}
    \phi(x) = \left(\begin{array}{c}
        \phi_c(x) \\ f(r) e^{in\theta} + \delta\phi(x)
    \end{array}\right), \quad
    \vec{\tilde{Z}} = -\frac{nz(r)}{r}\vec{e}_\theta + \delta\vec{\tilde{Z}}(x).
\end{align}
Here, $G_\mu^{\alpha}(x), \, G_\mu^{\bar{a}}(x)$ and $\tilde{A}_\mu(x)$ are the components of the gauge fields. 
Because the generalized Z-string solutions are independent of the $t$ and $z$-coordinates, the perturbation modes that depend on $t$ or $z$ can only make a positive contribution to the energy. This logic holds for the perturbation modes of the $t$ and $z$ components of the gauge fields. Hence we ignore them and discuss a variation of the energy linear density along the $z$-axis as in  the case of Z-string.

Since the generalized Z-string solutions are static and classical solutions, variational terms of the first order of the perturbation modes vanish and terms of the second order become leading. Thus we evaluate the sign of the quadratic terms in the energy linear density to check the stability. 
The variation of the energy linear density are divided into three parts by the transformation properties of perturbation modes under $SU(N-1)$ as $\mu=\mu_{\rm ad}+\mu_f+\mu_s$,
where $\mu_{\rm ad}$ includes only the adjoint representation modes $G^{\alpha}_\mu$,  $\mu_f$ includes the fundamental and anti-fundamental representation modes $\phi_c$ and $G^{\bar{a}}_\mu$, and $\mu_s$ includes the singlet perturbation modes $\delta\phi$, $\delta\tilde{Z}_\mu$, $\tilde{A}_\mu$. This is because the energy linear density is $SU(N-1)$ invariant.  $\mu_{\rm ad}$, $\mu_f$, and $\mu_s$ are explicitly written as follows.
\begin{align}
    \label{mu_ad}
    \mu_{\mbox{ad}} = \sum_{\alpha=1}^{(N-1)^2-1} \int d^2x \left[\frac{1}{2} \left(\nabla\times \vec{G}^{\alpha}\right)^2\right].
\end{align}
\begin{widetext}
\begin{eqnarray}
    \label{mu_f}
    \mu_f = 
    &\sum_{k=1}^{N-1} \int d^2x \left[ 2\left\{ \left(\nabla\times\vec{G}^{k+}\right)_z -\frac{im_G^2}{vm_{\tilde{Z}}}\frac{nz}{r}G^{k+}_r \right\} \left\{ \left(\nabla\times\vec{G}^{k-}\right)_z + \frac{im_G^2}{vm_{\tilde{Z}}}\frac{nz}{r}G^{k-}_r \right\} + \frac{2im_G^2}{vm_{\tilde{Z}}} \frac{nz'}{r} \left(\vec{G}^{k+}\times\vec{G}^{k-}\right)_z \right.\nonumber\\
    &\hspace{20mm} + \left| \partial_r \phi_{c,k} \right|^2 + \frac{1}{r^2}\left| \left(\partial_\theta - i\frac{m_{\tilde{Z}}}{2v} \left(1-\frac{2m_G^2}{m^2_{\tilde{Z}}}\right) nz\right) \phi_{c,k} \right|^2 + \frac{m_G^2}{v^2}f^2 G^{k+}_{\bar{i}}G^{k-}_{\bar{i}} + \frac{m_\phi^2}{4v^2}\left(f^2 -v^2\right)|\phi_{c,k}|^2 \nonumber\\
    &\hspace{20mm} -i\frac{m_G}{v} \left\{ \left(f' G^{k-}_r -if \frac{n(1-\frac{m_{\tilde{Z}}}{2v}z)}{r} G^{k-}_\theta\right)e^{-in\theta} \phi_{c,k} \right.\nonumber\\
    &\hspace{30mm} \left. + f e^{in\theta} \left(G^{k+}_r \partial_r\phi_{c,k}^* + \frac{G^{k+}_\theta}{r}\left(\partial_\theta + i\frac{m_{\tilde{Z}}}{2v}\left(1-\frac{2m_G^2}{m_{\tilde{Z}}^2}\right)nz\right)\phi_{c,k}^*\right) \right\} \nonumber\\
    &\hspace{20mm} +i\frac{m_G}{v} \left\{ \left(f' G^{k+}_r +if \frac{n(1-\frac{m_{\tilde{Z}}}{2v}z)}{r} G^{k+}_\theta\right)e^{in\theta} \left( \phi_{c,k}\right)^* \right.\nonumber\\
    &\hspace{30mm}\left. \left. + f e^{-in\theta} \left(G^{k-}_r \partial_r\phi_{c,k} + \frac{G^{k-}_\theta}{r}\left(\partial_\theta - i\frac{m_{\tilde{Z}}}{2v}\left(1-\frac{2m_G^2}{m_{\tilde{Z}}^2}\right)nz\right)\phi_{c,k}\right)  \right\} \right],
\end{eqnarray}
\end{widetext}
where
we write the components of $\phi_c$ and $G^{\bar{a}}_\mu$ as
\begin{align}
    \phi_c \equiv \left(\begin{array}{c}
        \phi_{c,1} \\ \vdots \\ \phi_{c,N-1}
    \end{array}\right) 
\end{align}
\begin{align}
    \left(\begin{array}{c}
        \vec{G}^{k+} \\ \vec{G}^{k-}
    \end{array}\right)
    \equiv \frac{1}{2}
    \left(\begin{array}{ccc}
        1 && -i \\ 1 && i
    \end{array}\right)
    \left(\begin{array}{c}
        \vec{G}^{(N-1)^2 + 2(k-1)} \\ \vec{G}^{(N-1)^2 + 2k-1}
    \end{array}\right) ,
\end{align}
where $k=1,\dots,N-1$.
\begin{align}
    \mu_s = \int d^2x &\left[
    \frac{1}{2}\left(\nabla\times \vec{\tilde{Z}}\right)^2 + \frac{1}{2}\left(\nabla\times \vec{\tilde{A}}\right)^2\right. \nonumber\\
    &\left.+\left| \left(\partial_{\bar{i}} + i\frac{m_{\tilde{Z}}}{2v} \tilde{Z}_{\bar{i}}\right) \phi_n \right|^2 + \frac{m_\phi^2}{8v^2}\left(|\phi_n|^2 -v^2\right)^2 \right] 
\end{align}
where $\phi_n(x)\equiv f(r) e^{in\theta} + \delta\phi(x)$ and $\vec{\tilde{Z}}(x) = -\frac{nz(r)}{r}\vec{e}_\theta + \delta\vec{\tilde{Z}}(x)$. 

The integrand in $\mu_{\mbox{ad}}$ is non-negative. In addition, $\mu_s$ is never smaller than the energy linear density of the generalized Z-string because it takes the same form as the one of the N-O string solution, which is classically stable,  in the $U(1)$ Higgs model. Therefore we will ignore them. As can be seen from Eq.(\ref{mu_f}), $\mu_f$ can be written as the sum of $N-1$ parts, $\mu_f \equiv \sum_k \mu_k$. All $\mu_k$ have the same form as the functional of $(\phi_{c,k}(x), G^{k\pm}_\mu(x))$ and thus it is sufficient to consider any one of them, for example, $\mu_1$.

If we non-dimentionalize $f(r), z(r)$ and $r$ as in Eq.(\ref{normalize_SUN}), $\mu_k$ becomes 
\begin{widetext}
\begin{align}
    \label{mu_k_mass_nondim}
    \mu_k = 
    &\int RdRd\theta \left[ 2\left\{ \left(\nabla\times\vec{G}^{k+}\right)_z -2i\frac{m_G^2}{m^2_{\tilde{Z}}}\frac{nZ}{R}G^{k+}_R \right\} \left\{ \left(\nabla\times\vec{G}^{k-}\right)_z + 2i\frac{m_G^2}{m^2_{\tilde{Z}}}\frac{nZ}{R}G^{k-}_R \right\} \right.\nonumber\\
    &\hspace{20mm} + 4i\frac{m_G^2}{m^2_{\tilde{Z}}} \frac{nZ'}{R} \left(\vec{G}^{k+}\times\vec{G}^{k-}\right)_z \nonumber\\
    &\hspace{20mm} + \left| \partial_R \phi_{c,k} \right|^2 + \frac{1}{R^2}\left| \left(\partial_\theta - i \left(1-\frac{2m_G^2}{m^2_{\tilde{Z}}}\right) nZ\right) \phi_{c,k} \right|^2 \nonumber\\
    &\hspace{20mm} -i\frac{2m_G}{m_{\tilde{Z}}} \left\{ \left(F' G^{k-}_R -iF \frac{n(1-Z)}{R} G^{k-}_\theta\right)e^{-in\theta} \phi_{c,k} \right.\nonumber\\
    &\hspace{20mm} \left. + F e^{in\theta} \left(G^{k+}_R \partial_R\phi_{c,k}^* + \frac{G^{k+}_\theta}{R}\left(\partial_\theta + i\left(1-\frac{2m_G^2}{m_{\tilde{Z}}^2}\right)nZ\right)\phi_{c,k}^*\right) \right\} \nonumber\\
    &\hspace{20mm} +i\frac{2m_G}{m_{\tilde{Z}}} \left\{ \left(F' G^{k+}_R +iF \frac{n(1-Z)}{R} G^{k+}_\theta\right)e^{in\theta} \left( \phi_{c,k}\right)^* \right.\nonumber\\
    &\hspace{20mm} \left. + F e^{-in\theta} \left(G^{k-}_R \partial_R\phi_{c,k} + \frac{G^{k-}_\theta}{R}\left(\partial_\theta - i\left(1-\frac{2m_G^2}{m_{\tilde{Z}}^2}\right)nZ\right)\phi_{c,k}\right)  \right\} \nonumber\\
    &\hspace{20mm} \left.+ \frac{4m_G^2}{m_{\tilde{Z}}^2}F^2 G^{k+}_{\bar{i}}G^{k-}_{\bar{i}} + \frac{m_\phi^2}{m_{\tilde{Z}}^2}\left(F^2 -1\right)|\phi_{c,k}|^2 \right]
\end{align}
\end{widetext}
Note that 
$\mu_k$ for any perturbation modes $\vec{G}^{k\pm}(x)$ and $\phi_{c,k}(x)$ is determined by $F(R)$, $Z(R)$, and two mass ratios $m_{\phi}/m_{\tilde{Z}}$ and $m_G/m_{\tilde{Z}}$.
Since $F(R)$ and $Z(R)$ are determined by $m_{\phi}/m_{\tilde{Z}}$, the classical stability of the generalized Z-string is determined only
by two mass ratios $m_{\phi}/m_{\tilde{Z}}$ and $m_G/m_{\tilde{Z}}$.  It is important that 
it does not explicitly depend on $N$ in Eq. (\ref{mu_k_mass_nondim}). Since the region for the stability of the string solution for $N=2$ has already been shown in Ref.\cite{James:1992zp}, the regions for the stability of the generalized Z-string solution in parameter 
space of the above two mass ratios
can be understood by replacing
these boson masses for $N=2$ with those for any $N$.\footnote{Note that if we use other parameters instead of these mass ratios, for example, the mass ratio $m_\phi/m_{\tilde Z}$ and the mixing $\sin\theta_G$ as in the case of Z-string\cite{James:1992zp}, the stability depends explicitly on $N$, that can be seen in the appendix. Therefore, the argument here becomes more difficult for other parameterizations. %However, of course, we have obtained the same results even by more steady calculation, which is given  in the appendix.
} 

\begin{figure}[t]
    \centering
    \includegraphics[keepaspectratio, scale=0.6]{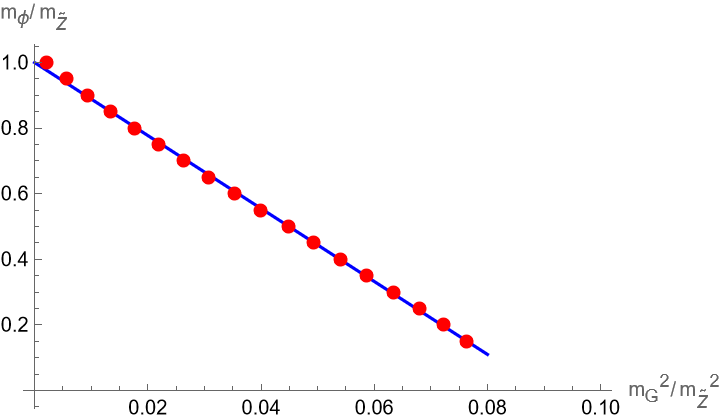}
    \caption{In this figure, we have shown the region in which the generalized Z-string becomes classically stable, in the parameter space of  $m_G^2/m_{\tilde{Z}}^2$ and $m_{\phi}/m_{\tilde{Z}}$. The region to the left of the red dots, which are obtained by our numerical calculation,  indicates the region where the generalized Z-string is stable. 
%that each of them becomes the boundary of whether the generalized Z-string is classically stable or not for each . The set of values have been shown as red points in the parameter space of the two mass ratios. 
The blue straight line through $(m_G^2/m_{\tilde{Z}}^2, m_{\phi}/m_{\tilde{Z}})=(0,1)$ is obtained by using the least squared method, and the concrete formula is shown in Eq.(\ref{stable_estimation_N}). 
%We can estimate the generalized Z-string is classically stable in the left region of the blue line.
}
    \label{gZstring_stableregion}
\end{figure}

We have shown our numerical result of the stability analysis for the generalized Z-string in the Fig. \ref{gZstring_stableregion} as 
a plot in the parameter space of the two mass ratios 
$(m_{\phi}/m_{\tilde{Z}}, m_G^2/m_{\tilde{Z}}^2)$. The numerical methods are explained below. First, for fixed $m_{\phi}/m_{\tilde{Z}}$,
we calculate the generalized Z-string solution. In more detail, two dimensionless functions $F(R)$ and $Z(R)$  with the range $10^{-11}\leq R\leq 50$ are obtained by %solving theequations of motion by 
the gradient flow of the energy whose density have been shown in Eq.(\ref{e_density}). Second, we check the stability of the solutions for fixed
two mass ratios $m_{\phi}/m_{\tilde{Z}}$ and $m_G^2/m_{\tilde{Z}}^2$.
 Concretely, we calculate the sign of the minimum eigenvalue of the operator $\tilde{\mathcal{O}}_N$ in Eq.(\ref{O_gZstring}) in the appendix by Mathematica\cite{mathematica}. 
Finally, for fixed $m_{\phi}/m_{\tilde{Z}}$, we obtain the maximal value of $m_G^2/m_{\tilde{Z}}^2$ for the stable generalized Z-string.  It has been checked that changing the maximum value of $R$ does not change the result. 
If $g_2=0$ (i.e., $m_G^2/m_{\tilde{Z}}^2=0$) and $N=2$, the solution is known as the semi-local string\cite{Vachaspati:1991dz}, which is stable when  $m_{\phi}/m_{\tilde{Z}}\leq 1$ as in Ref.\cite{Hindmarsh:1991jq}. 
On the other hand, our obtained maximal value of $m_G^2/m_{\tilde{Z}}^2$ for $m_{\phi}/m_{\tilde{Z}}=1$ is 0.0021. It means that there is at least an error of $\mathcal{O}(0.001)$ in our calculation. We think that our results are qualitatively consistent with the results in Ref.\cite{James:1992zp}, although the quantitative consistency is not clear because the errors of both calculations are not clear.  %Therefore, our calculation We have calculate the 
%is smaller than $m_G^2/m_{\tilde{Z}}^2$ is calculated in Ref. []. The result

%$m_G^2/m_{\tilde{Z}}^2$ for fixed $m_{\phi}/m_{\tilde{Z}}$
%so that the generalized Z-string is stable 

Since the data points in Fig. \ref{gZstring_stableregion} are approximately on a straight
line, we can obtain a linear approximate formula for the region where the generalized Z-string becomes classically stable as 
\begin{align}
    \label{stable_estimation_N}
    \frac{m_{\phi}}{m_{\tilde{Z}}} \lesssim 1 - 11 \frac{m_G^2}{m_{\tilde{Z}}^2} 
\end{align}
by using the least squared method. Note that this formula is not reliable for 
$\frac{m_{\phi}}{m_{\tilde{Z}}}<0.15$, where the numerical calculation becomes more difficult.
We will apply this approximate formula to several patterns of symmetry breaking later.

%%%%%%%%%%%%%%%%%%%%%%%%%%%%%%%
\section{Embedded string in SUSY $SU(N)\times U(1)_X$ model}
%%%%%%%%%%%%%%%%%%%%%%%%%%%%%%%

In this section, we consider about the embedded string in the SUSY $SU(N)\times U(1)_X$ gauge theory with two Higgses whose VEVs break $SU(N)\times U(1)_X$ into $SU(N-1)\times U(1)_Q$. It is a SUSY extension of the generalized Z-string. We will show that its classical stability also can be determined by the ratios of the masses of the Higgs and the massive gauge bosons. We will find that the stability condition for the solution becomes the same as that in the non-SUSY model, while the stability region cannot increase but may decrease in general when the number of Higgses increase. The case of $N=2$ has already been studied in Ref.\cite{Kanda:2022xrz}. 

We consider a SUSY $SU(N)\times U(1)_X$ gauge theory in which  a $SU(N)$ fundamental representation Higgs $\Phi_1$ and an anti-fundamental representation Higgs $\Phi_2$ are introduced  as chiral superfields. Both $\Phi_1$ and $\Phi_2$ have $U(1)_X$ charges $1/2$ and $-1/2$, respectively. The superpotential of the Higgses are given as
\begin{align}
    W = \lambda_s S \left(\Phi_2^\top \Phi_1 - u^2\right) \qquad (\lambda_s, u \in \mathbb{R}) ,
\end{align}
where $S$ is a gauge singlet chiral superfield.  Here, without loss of generality, the parameters $\lambda_s$ and $u$ can be taken real.

We write the scalar component fields  $\Phi_1$, $\Phi_2$ and $S$ as $\phi_1$, $\phi_2$ and $s$, respectively. 
The F-term contributions are given as
\begin{align}
    \label{F_contribution}
    V_F (\phi_1, \phi_2, s) = \lambda_s^2 \left| \phi_2^\top\phi_1 - u^2 \right|^2 + \lambda_s^2 \left|s\right|^2 \left(\left|\phi_1\right|^2 + \left|\phi_2\right|^2\right) ,
\end{align}
and the D-term contribution is also given as
\begin{align}
    \label{D_contribution}
    V_D (\phi_1, \phi_2) &= \frac{g_1^2}{8} \left(\left|\phi_1\right|^2 - \left|\phi_2\right|^2\right)^2 \nonumber\\
    &\hspace{3mm}+ \frac{g_N^2}{2} \left(\phi_1^\dagger T^a \phi_1 - \phi_2^\top T^a \phi_2^*\right) ^2 \nonumber\\
    &= \left(\frac{g_1^2}{8} - \frac{g_N^2}{4N}\right) \left(\left|\phi_1\right|^2 - \left|\phi_2\right|^2\right)^2 \nonumber\\
    &\hspace{3mm}+ \frac{g_N^2}{4} \left( \left|\phi_1\right|^4 + \left|\phi_2\right|^4 - 2 \left|\phi_2^\top\phi_1\right|^2 \right) .
\end{align}
In the last step in Eq.(\ref{D_contribution}), we have used a Fierz identity. The F-flatness conditions are given as
\begin{align}
    \phi_2^\top\phi_1 = u^2, \quad s\phi_1 = s\phi_2 = 0,
\end{align}
and the D-flatness conditions become
\begin{align}
    \left|\phi_1\right|^2 - \left|\phi_2\right|^2 = \phi_1^\dagger T^a \phi_1 - \phi_2^\top T^a \phi_2^* = 0 .
\end{align}
Therefore, we can find that the VEVs  
are %given as
\begin{align}
    \label{susy_vev}
    \phi_1 = \left(\begin{array}{c}
        0 \\ \vdots \\ 0 \\ u
    \end{array}\right) , \quad
    \phi_2 = \left(\begin{array}{c}
        0 \\ \vdots \\ 0 \\ u
    \end{array}\right) , \quad
    s = 0 , 
\end{align}
which break $SU(N)\times U(1)_X$ gauge symmetries into $SU(N-1)\times U(1)_Q$. 

The ansatz for the embedded string solutions can be written as 
\begin{align}
    \label{susy_gZstring}
    \begin{aligned}
        &\phi_1 = \left(\begin{array}{c}
            0 \\ \vdots \\ 0 \\ f_1(r) e^{in\theta}
        \end{array}\right) , \quad
        \phi_2 = \left(\begin{array}{c}
            0 \\ \vdots \\ 0 \\ f_2(r) e^{-in\theta}
        \end{array}\right) , \\
        &\tilde{Z}_\theta(x) = - nz(r) , \qquad (n\in \mathbb{Z}\setminus \{0\})\\
         &\tilde{Z}_t(x) = \tilde{Z}_r(x) = \tilde{Z}_z(x) = \tilde{A}_\mu(x) = G^{\alpha}_\mu = G^{\bar{a}}_\mu =0 ,
    \end{aligned}
\end{align}
where $f_1(r)$, $f_2(r)$ and $z(r)$ are real functions. The winding numbers of $\phi_1$ and $\phi_2$ are the same absolute values and opposite signs because of the F-flatness conditions when $r\rightarrow\infty$. $f_1(r)$, $f_2(r)$ and $z(r)$ have to satisfy the boundary conditions which are given as 
\begin{align}
    &f_1(0) = f_2(0) = z(0) = 0 , \nonumber\\
    &f_1(\infty) = f_2(\infty) = u , \quad z(\infty) = \frac{2}{\alpha_N} ,
\end{align}
because they have to be single valued when $r=0$ and make the energy density zero when $r\rightarrow\infty$. Furthermore, we can conclude that $f_1(r)=f_2(r)$  because the action is invariant to the exchange of $f_1(r)$ and $f_2(r)$ and they have the same boundary conditions. Actually, this conclusion can be numerically confirmed.
Hereinafter, we write $f_1(r)=f_2(r)\equiv f(r)$. The equation of motions for $f(r)$ and $z(r)$ become 
\begin{align}
    \label{feq_susySUN}
    &f'' + \frac{f'}{r} - n^2\left(1-\frac{\alpha_N}{2}z\right) \frac{f}{r^2} + \lambda_s^2 \left(u^2 - f^2\right) f = 0 \\
    \label{zeq_susySUN}
    &z'' - \frac{z'}{r} + 2\alpha_N \left(1-\frac{\alpha_N}{2}z\right) f^2 =0 .
\end{align}

We examine to see whether the classical solutions are stable or not by perturbation method as in the previous sections.  
The perturbations from Eq.(\ref{susy_gZstring}) can be written as
\begin{align}
    \begin{aligned}
        &\phi_1 = \left(\begin{array}{c}
            \phi_{1c,1}(x) \\ \vdots \\ \phi_{1c,N-1}(x) \\ f(r) e^{in\theta} + \delta\phi_1(x)
        \end{array}\right) , \quad \\
        &\phi_2 = \left(\begin{array}{c}
            \phi_{2c,1}(x) \\ \vdots \\ \phi_{2c,N-1}(x) \\ f(r) e^{-in\theta} + \delta\phi_2(x)
        \end{array}\right) , \\
        &\vec{\tilde{Z}}(x) = - \frac{nz(r)}{r} \vec{e}_\theta + \delta\vec{\tilde{Z}}(x) ,
    \end{aligned}
\end{align}
and we also consider
\begin{align}
    G^{\alpha}_\mu(x), \quad G^{\bar{a}}_\mu(x), \quad \tilde{A}_\mu(x) \quad \mbox{and}\quad s(x),
\end{align}
as perturbations. For the same reasons as in the non-SUSY case, we can ignore $t$ and $z$-coordinates dependence and the $t$ and $z$ components of the gauge fields. Furthermore, as in the previous section that the variation of the energy linear density can be divided into three categories in terms of $SU(N-1)$ representation of the perturbation modes. 
The part which includes $G^{\alpha}_\mu(x)$ (adjoint representation) has the same form as in Eq.(\ref{mu_ad}) and it is non-negative. 
The part which includes singlets, $\delta\phi_1(x), \delta\phi_2(x), \delta\vec{\tilde{Z}}(x)$, and $\vec{\tilde{A}}(x)$ are given as
\begin{align}
    \label{susyvariation_n}
    \int rdrd\theta &\left[ \frac{1}{2} \left(\nabla\times\vec{\tilde{Z}}\right)^2 + \left| \left(\partial_{\bar{i}} + i\frac{\alpha_N}{2}\tilde{Z}_{\bar{i}}\right) \phi_{1n} \right|^2 \right. \nonumber\\
    & + \left| \left(\partial_{\bar{i}} - i\frac{\alpha_N}{2}\tilde{Z}_{\bar{i}}\right) \phi_{2n} \right|^2 + \left| \partial_{\bar{i}} s \right|^2 \nonumber\\
    & + \lambda_s^2 \left| \phi_{1n} \phi_{2n} - u^2 \right|^2 
    + \lambda_s^2 |s|^2 \left(\left| \phi_{1n} \right|^2 + \left| \phi_{2n} \right|^2 \right) \nonumber\\
    &\left.+ \frac{\alpha_N^2}{8} \left(\left| \phi_{1n} \right|^2 - \left| \phi_{2n} \right|^2 \right)^2 + \frac{1}{2} \left(\nabla\times\vec{\tilde{A}}\right)^2 \right] , 
\end{align}
where $\phi_{1n}(x) \equiv f(r) e^{in\theta} + \delta\phi_1(x)$ and $\phi_{2n}(x) \equiv f(r) e^{-in\theta} + \delta\phi_2(x)$. In the integrand of Eq.(\ref{susyvariation_n}), the fourth, sixth and eighth terms must be non-negative and we can set $s=0$ and $\nabla\times\vec{\tilde{A}}=0$, which make these terms vanishing. Thus, Eq.(\ref{susyvariation_n}) becomes the same energy linear density as that of $U(1)$ gauge theory with two Higgses, 
$\phi_{1n}$ and $\phi_{2n}$ which is non-negative 
because the N-O string with two Higgses is also a topological defect\cite{La:1993je}. 

Only perturbations of the $SU(N-1)$ fundamental (and anti-fundamental) part of $\phi_1$ and $\phi_2$, and $G^{\bar{a}}_\mu$ can make a variation of the energy density negative. 
We call them the charged perturbations in the following.
We will show that the arguments for classically stable string solution become the same as those in non-SUSY case. 
Let us rotate  $\phi_1$ and the complex conjugate of $\phi_2$ as
\begin{align}
    \left(\begin{array}{c}
        \phi_0(x) \\ \phi(x)
    \end{array}\right)
    \equiv \frac{1}{\sqrt{2}}\left(\begin{array}{cc}
        1 & -1 \\ 1 & 1
    \end{array}\right)
    \left(\begin{array}{c}
        \phi_1(x) \\ \phi_2^*(x)
    \end{array}\right) ,
\end{align}
where the $\phi_0(x)$ has a zero VEV, $\braket{\phi_0}=0$.
If we write the charged perturbations as\footnote{Note that Higgs part of the embedded string solutions in Eq.(\ref{susy_gZstring}) are shown as $\phi=\sqrt{2}f(r)e^{in\theta}$.}
\begin{align}
    \phi_0(x) = \left(\begin{array}{c}
        \phi_{0c} (x) \\ 0
    \end{array}\right), \quad
    \phi(x) = \left(\begin{array}{c}
        \phi_c(x) \\ \sqrt{2}f(r) e^{in\theta}
    \end{array}\right) ,
\end{align}
the variation of the Higgs potential $V_F$ in Eq.(\ref{F_contribution}) and $V_D$ in Eq.(\ref{D_contribution}) are given as
\begin{align}
    \label{susy_charged_variation}
    \delta V = \left[\lambda_s^2\left(u^2 - f^2\right) + g_N^2 f^2 \right] \left|\phi_{0c}\right|^2 - \lambda_s^2 \left(u^2 - f^2\right) \left|\phi_c\right|^2 .
\end{align}
The first term in Eq.(\ref{susy_charged_variation}) does not become negative because $f(r)$ is a real function satisfying $0\leq f(r) < u$. The variation of the gradient energy of $\phi_{0c}$ also must be non-negative, thus the perturbation of $\phi_{0}$ does not affect the stability of the embedded string. 

If we rewrite the Higgs potential $V(\phi_0, \phi, s)\equiv V_F+V_D$ without $\phi_0$, we obtain that
\begin{align}
    \label{susy_final_potential}
    V(\phi_0=0, \phi, s) = \frac{\lambda_s^2}{4} \left(\left|\phi\right|^2 - 2u^2\right)^2 + \lambda_s^2 \left|s\right|^2 \left|\phi\right|^2 .
\end{align}
If we ignore the second term in Eq.(\ref{susy_final_potential}) since it does not affect the stability, we find that it is nothing but the Mexican hat potential which is the same as the potential in Sec.3 by replacing $\lambda$ and $v$ with $\frac{\lambda_s^2}{4}$ and $\sqrt{2}u$, respectively. Thus we can conclude that the classical stability of the generalized Z-string in the SUSY $SU(N)\times U(1)_X$ Higgs model is determined by $m_\phi/m_{\tilde{Z}}$ and $m_G/m_{\tilde{Z}}$, where $m_\phi, m_{\tilde{Z}}$ and $m_G$ are the masses of the neutral part of $\phi$, the neutral massive gauge boson and the charged massive gauge boson, respectively. The region in the parameter space of $(m_\phi/m_{\tilde{Z}}, m_G/m_{\tilde{Z}})$ where the embedded string is classically stable is the same as the region in Fig. \ref{gZstring_stableregion} that we derived in Sec.3.

%%%%%%%%%%%%%%%%%%%%%%%%%%%%%%%%
\section{Applications}
%%%%%%%%%%%%%%%%%%%%%%%%%%%%%%%%
In this section, we will consider the application of the stability conditions for generalized
Z-string.  We would like to clarify which models can produce the stable generalized Z-string. First of all, the stability condition in Eq.(\ref{stable_estimation_N}) is rewritten to the
condition between gauge couplings, because this form of the condition is easier to be applied to concrete models than the condition (\ref{stable_estimation_N}).
It is given as
\begin{align}
    \label{constraint_couplings}
    g_1 \gtrsim \sqrt{\frac{11}{1-m_\phi/m_{\tilde{Z}}} - \frac{2(N-1)}{N}}\, g_N ,
\end{align}
where the $U(1)_X$ gauge coupling is normalized so that the charge of Higgs is taken as
1/2. 
For example, if we take $m_\phi\ll m_{\tilde Z}$ and $N\rightarrow\infty$, we obtain $g_1>3g_N$.
Since this is almost the minimum lower bound for $g_1$, 
we can conclude that very large $g_1$ is needed  to obtain the stable string.
The stable generalized Z-string can be produced
when this condition (\ref{constraint_couplings}) is satisfied at the phase transition.

Next, we consider situations in which the gauge couplings of $SU(N)$ and $U(1)_X$ are related with each other. 
Suppose that $SU(N)$ and $U(1)_X$ gauge interactions are unified into $SU(N+1)$ gauge interaction. 
The $SU(N+1)$ fundamental representation is divided with  the $SU(N)\times U(1)_X$ representation as
\begin{align}
    \label{SUN1_U1_normaliz}
    \mathbf{N+1} = \mathbf{N}_c \oplus \mathbf{1}_{-Nc} ,
\end{align} 
where the indices denote the $U(1)_X$ charges normalized so that $g_1=g_N$\footnote{Here we do not consider the renormalization group effects.} , that results in $c=1/\sqrt{2N(N+1)}$.
When the $SU(N)$ fundamental Higgs has the $U(1)_X$ charge $q$, the Eq.(\ref{constraint_couplings}) can be rewritten as 
\begin{align}
    \label{constraint_charge_general}
    q^2 \gtrsim \frac{2.75}{1-m_\phi/m_{\tilde{Z}}} - \frac{N-1}{2N} .
\end{align}
This condition (\ref{constraint_charge_general}) gives the lower bound for $q$, which
must be quite large. 

\begin{figure}[t]
    \centering
    \includegraphics[keepaspectratio, scale=0.5]{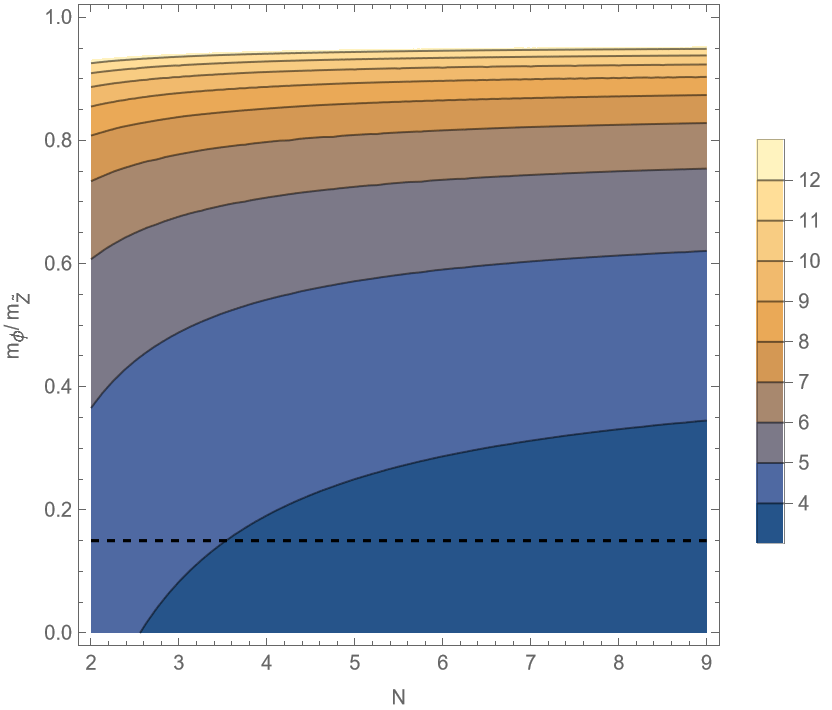}
    \caption{If the Higgs is the $SU(N)$ fundamental representation included in the completely symmetric $k$ th-rank tensor representation of $SU(N+1)$, the minimum value of $k$ for the classically stable Z-string is shown. Each color in the graph corresponds to the minimum value of $k$. The dashed line denotes $m_\phi/m_{\tilde Z}=0.15$, which is the minimum value that we have checked in our numerical calculation.}
    \label{SUN1_figure}
\end{figure}

Let us consider how to realize such a large $q$. 
If we consider  a completely symmetric $k$ th-rank tensor representation field of $SU(N+1)$, it includes an $SU(N)$ fundamental representation field with the $U(1)_X$ charge $q=(1-(k-1)N)/\sqrt{2N(N+1)}$. 
Thus the Eq.
(\ref{constraint_charge_general}) becomes 
\begin{align}
    \label{constraint_SUN1}
    k^2 - \frac{2(N+1)}{N} k - \frac{N+1}{N}\left(\frac{5.5}{1-m_\phi/m_{\tilde{Z}}}-2\right) \gtrsim 0  .
\end{align}
For given $N$ and $m_\phi/m_{\tilde Z}$, the minimum value of $k$ to stabilize the generalized Z-string classically can be obtained by Eq.(\ref{constraint_SUN1}) as shown in Fig.\ref{SUN1_figure}. From the Fig. \ref{SUN1_figure}, we find the region 
in which  the generalized Z-string can be stabilized if we take $k\geq 4$.

In practice, the renormalization group effects must be considered. 
When we take $g_N = \alpha_{RG} g_1 \, (\alpha_{RG}>0)$, where $\alpha_{RG}$ denotes the renormalization group effect, 
the Eq.(\ref{constraint_charge_general}) becomes
\begin{align}
    \label{constraint_charge_RGE}
    q^2 \gtrsim \alpha_{RG}^2 \left[\frac{2.75}{1-m_\phi/m_{\tilde{Z}}} - \frac{N-1}{2N}\right] .
\end{align}
Since $\alpha_{RG}$ is usually larger than 1, $|q|$ must be larger. However, in principle, it is 
possible that $\alpha_{RG}<1$ as discussed later. 
Thus, the minimum value of $|q|$ and $k$ for the classically stable Z-string become smaller than in the above example. 

The above discussion can be extended to the general unified group $G$. 
As seen in the above, 
higher rank representation field of $G$ is important to obtain the $SU(N)$ fundamental Higgs with large $q$. In the followings, we consider two famous patterns of  symmetry breaking from $SO(10)$ to the SM gauge group and discuss what happens when we apply the condition for the classically stable generalized Z-string to these scenarios. 

As a first example, we consider a case where $SO(10)\rightarrow SU(3)_C\times SU(2)_L\times SU(2)_R\times U(1)_X\rightarrow SU(3)_C\times SU(2)_L\times U(1)_Y$.
The first symmetry breaking can be caused by developing a VEV of adjoint Higgs of $SO(10)$. 
If the second symmetry breaking is caused by Higgs $(\mathbf{1},\mathbf{1},\mathbf{2},q)\,(q\neq 0)$ under $SU(3)_C\times SU(2)_L\times SU(2)_R\times U(1)_X$,  Eq.(\ref{constraint_charge_RGE}) with $N=2$ can be applied for the stability of the generalized Z-string as
\begin{align}
    \label{constraint_charge_so10ex1}
    q^2 \gtrsim \alpha_{RG}^2 \left[\frac{2.75}{1-m_\phi/m_{\tilde{Z}}} - \frac{1}{4}\right] ,
\end{align}
where $U(1)_X$ is normalized so that the gauge coupling constants of $SU(3)_C, SU(2)_L, SU(2)_R$, and $U(1)_X$ are equal at the unification scale. If the Higgs doublet is from 
$\mathbf{16}$ of $SO(10)$, $q$ becomes $\frac{3}{2\sqrt{6}}$, which is too small to satisfy the Eq. (\ref{constraint_charge_so10ex1}) as addressed in Refs. \cite{Kanda:2022xrz,Holman:1992rv,Buchmuller:2021dtt}. Here, we consider other cases where the
Higgs comes from higher dimensional representations. Such Higgses can have larger $U(1)_X$ charge $q=\pm \frac{3(2l+1)}{2\sqrt{6}}, (l=1,2,3,\cdots)$ and 
the higher dimensional representations of $SO(10)$ mentioned in Ref.\cite{Yamatsu:2015npn} are written in Table \ref{higgsses_so10ex1}.

\begin{table*}[t]
    \caption{the representations of $SO(10)$ which include $\left(\mathbf{1},\mathbf{1},\mathbf{2},q\right)$ under $SU(3)_C\times SU(2)_L\times SU(2)_R\times U(1)_X$}
    \label{higgsses_so10ex1}
    \centering
    \begin{tabular}{|c|l|}
        \hline
        Included representaions & \multicolumn{1}{c|}{Representations of $SO(10)$} \\
        \hline\hline
        \multirow{4}{*}{$\left(\mathbf{1},\mathbf{1},\mathbf{2},\pm\frac{9}{2\sqrt{6}}\right)$ or $\left(\mathbf{1},\mathbf{1},\overline{\mathbf{2}},\pm\frac{9}{2\sqrt{6}}\right)$}
        &
        $\mathbf{1200},$ $\mathbf{8800},$ $\mathbf{11088},$ $\mathbf{17280},$ $\mathbf{25200},$ $\mathbf{30800},$ $\mathbf{34992},$ \\
        & $\mathbf{38016},$ $\mathbf{49280},$ $\mathbf{55440},$ $\mathbf{102960},$ $\mathbf{124800},$ $\mathbf{144144},$ \\
        & $\mathbf{164736},$ $\mathbf{258720},$ $\mathbf{428064},$ $\mathbf{465696}$, \\
        & and their complex conjugate representations (c.c.) \\
        \hline
        \multirow{2}{*}{$\left(\mathbf{1},\mathbf{1},\mathbf{2},\pm\frac{15}{2\sqrt{6}}\right)$ or $\left(\mathbf{1},\mathbf{1},\overline{\mathbf{2}},\pm\frac{15}{2\sqrt{6}}\right)$}
        &
        $\mathbf{30800},$ $\mathbf{196560},$ $\mathbf{364000},$ $\mathbf{428064},$ $\mathbf{465696},$ $\mathbf{498960}$, \\
        & and c.c. \\
        \hline
        $\left(\mathbf{1},\mathbf{1},\mathbf{2},\pm\frac{21}{2\sqrt{6}}\right)$ or $\left(\mathbf{1},\mathbf{1},\overline{\mathbf{2}},\pm\frac{21}{2\sqrt{6}}\right)$
        &
        $\mathbf{428064}$ and c.c. \\
        \hline
    \end{tabular}
\end{table*}

\begin{figure*}[t]
    \begin{minipage}[t]{0.5\linewidth}
        \centering
        \includegraphics[keepaspectratio, scale=0.3]{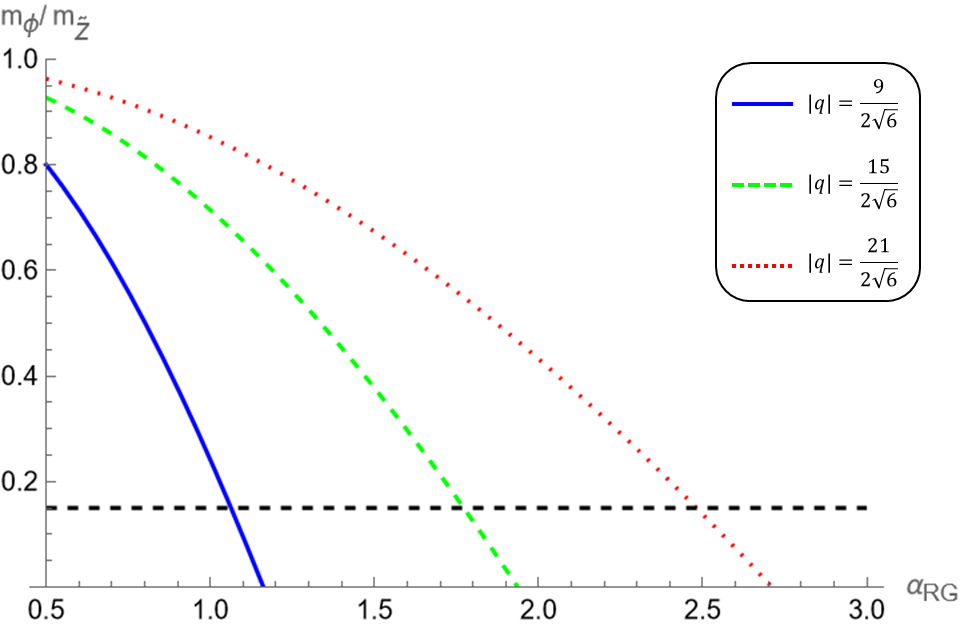}
    \end{minipage}%
    \begin{minipage}[t]{0.5\linewidth}
        \centering
        \includegraphics[keepaspectratio, scale=0.3]{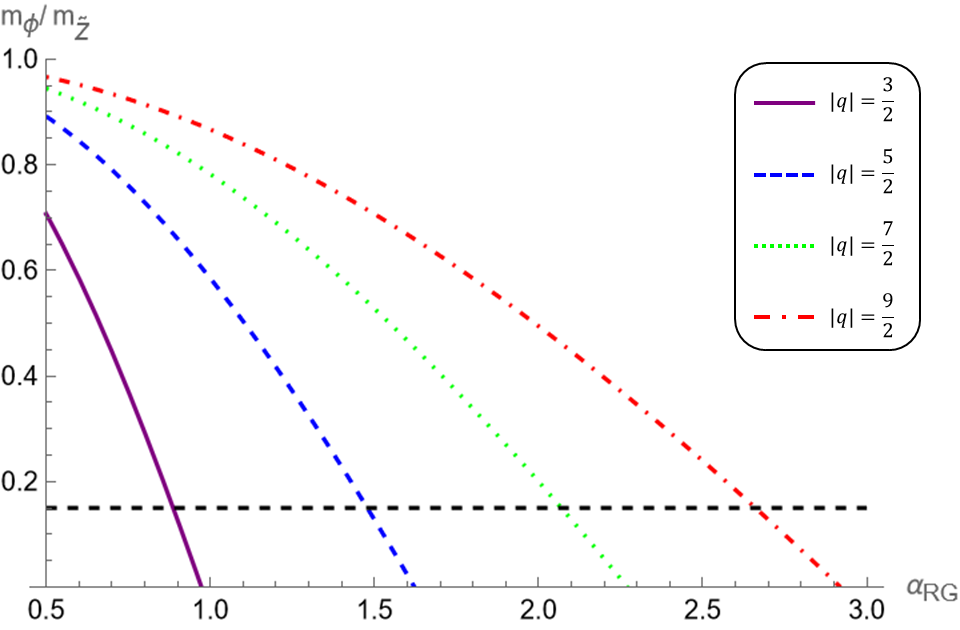}
    \end{minipage}
    \caption{The region in the parameter space of $(m_\phi/m_{\tilde{Z}}, \alpha_{RG})$ to make the generalized Z-string classically stable in the second symmetry breaking in $SO(10)\rightarrow SU(3)_C\times SU(2)_L\times SU(2)_R\times U(1)_X\rightarrow SU(3)_C\times SU(2)_L\times U(1)_Y$ (left) and $SO(10)\rightarrow SU(4)_C\times SU(2)_L\times U(1)_X\rightarrow SU(3)_C\times SU(2)\times U(1)_Y$ (right). The left regions of each lines are the regions where the generalized Z-string becomes classically stable for each case, respectively. The dashed lines denote $m_\phi/m_{\tilde{Z}}=0.15$, which is the minimum value that we have checked in our numerical calculation.
    }
    \label{Fig_ex1}
\end{figure*}

We show the regions in parameter space of $(m_\phi/m_{\tilde{Z}},\alpha_{RG})$ where the generalized Z-string becomes classically stable in the left graph in Fig.\ref{Fig_ex1}. 
The generalized Z-string becomes classically stable in the below region of each line for each $|q|$. We can see that the bigger $|q|$ gives the larger stability region. 
In the Fig. \ref{Fig_ex1}, we show the results for 
$\alpha_{RG}>0.5$, because $\alpha_{RG}$ can be smaller than 1
\footnote{
For example, in the breaking $SO(10)\rightarrow SU(4)_C\times SU(2)_L\times SU(2)_R\rightarrow SU(3)_C\times SU(2)_L\times SU(2)_R\times U(1)_X\rightarrow SU(3)_C\times SU(2)_L\times U(1)_Y$, the gauge coupling of $U(1)_X$ can be almost the same as $SU(3)_C$ which is expected to be larger than $SU(2)_R$. 
} 
although $\alpha_{RG}$ is usually  expected to be larger
than 1 due to non-Abelian property of $SU(2)_R$. 

Next, we consider another case  in which $SO(10)\rightarrow SU(4)_C\times SU(2)_L\times U(1)_X\rightarrow SU(3)_C\times SU(2)\times U(1)_Y$.
If the second symmetry breaking is caused by developing a VEV of Higgs whose representation is $(\mathbf{4},\mathbf{1},q)\,(q\neq 0)$ of $SU(4)_C\times SU(2)_L\times U(1)_X$, we can consider the generalized Z-string with $N=4$. We normalize the gauge coupling constants of $SU(4)_C$ and $U(1)_X$ so that they are equal at the unification scale. Thus we obtain the constraint for the $U(1)_X$ charge $q$ by Eq.(\ref{constraint_charge_RGE}) with $N=4$ as
\begin{align}
    \label{constraint_charge_so10ex2}
    q^2 \gtrsim \alpha_{RG}^2 \left[\frac{2.75}{1-m_\phi/m_{\tilde{Z}}} - \frac{3}{8}\right] ,
\end{align}
where $\alpha_{RG}$ is the ratio of the gauge coupling constant of $SU(4)_C$ to the one of $U(1)_X$ when $SU(4)_C\times U(1)_X$ are broken to $SU(3)_C\times U(1)_Y$. 
If the Higgs doublet is from 
$\mathbf{16}$ of $SO(10)$, $q$ becomes $\pm\frac{1}{2}$, which is too small to satisfy the Eq. (\ref{constraint_charge_so10ex2}).
Here, we consider other cases where the
Higgs comes from higher dimensional representations as in the first case. Such Higgses can have larger $U(1)_X$ charge $q=\pm \frac{2l+1}{2}, (l=1,2,3,\cdots)$ and 
the higher dimensional representations of $SO(10)$ mentioned in Ref.\cite{Yamatsu:2015npn} are written in Table \ref{higgsses_so10ex2}.
For the above each set of representations, we show the regions in parameter space of $(m_\phi/m_{\tilde{Z}},\alpha_{RG})$ where the generalized Z-string is classically stable in the right graph in Fig.\ref{Fig_ex1}. 

\begin{table*}[t]
    \caption{the representations of $SO(10)$ which include $\left(\mathbf{4},\mathbf{1},q\right)$ under $SU(4)_C\times SU(2)_L\times U(1)_X$}
    \label{higgsses_so10ex2}
    \centering
    \begin{tabular}{|c|l|}
        \hline
        Included representaions & \multicolumn{1}{c|}{Representations of $SO(10)$} \\
        \hline\hline
        \multirow{4}{*}{$\left(\mathbf{4},\mathbf{1},\pm\frac{3}{2}\right)$ or $\left(\overline{\mathbf{4}},\mathbf{1},\pm\frac{3}{2}\right)$} 
        &
        $\mathbf{560},$ $\mathbf{3696},$ $\mathbf{8064},$ $\mathbf{8800},$ $\mathbf{15120},$ $\mathbf{25200},$ $\mathbf{34992},$ $\mathbf{38016},$ $\mathbf{43680},$ \\
        & $\mathbf{48048},$ $\mathbf{70560},$ $\mathbf{124800},$ $\mathbf{129360},$ $\mathbf{144144},$ $\mathbf{155232},$ $\mathbf{196560},$ \\
        & $\mathbf{205920},$ $\mathbf{258720},$ $\mathbf{308880},$ $\mathbf{332640},$ $\mathbf{343200},$ $\mathbf{364000},$ $\mathbf{388080},$ \\
        & $\mathbf{443520},$ $\mathbf{443520'},$ $\mathbf{465696},$ $\mathbf{529200}$ , and c.c. \\
        \hline
        \multirow{2}{*}{$\left(\mathbf{4},\mathbf{1},\pm\frac{5}{2}\right)$ or $\left(\overline{\mathbf{4}},\mathbf{1},\pm\frac{5}{2}\right)$} 
        & 
        $\mathbf{8064},$ $\mathbf{43680},$ $\mathbf{70560},$ $\mathbf{144144},$ $\mathbf{155232},$ $\mathbf{258720},$ $\mathbf{332640},$ \\
        & $\mathbf{388080},$ $\mathbf{443520},$ $\mathbf{443520'},$ $\mathbf{529200}$ , and c.c. \\
        \hline
        $\left(\mathbf{4},\mathbf{1},\pm\frac{7}{2}\right)$ or $\left(\overline{\mathbf{4}},\mathbf{1},\pm\frac{7}{2}\right)$
        &
        $\mathbf{70560},$ $\mathbf{332640},$ $\mathbf{443520'},$ and c.c. \\
        \hline
        $\left(\mathbf{4},\mathbf{1},\pm\frac{9}{2}\right)$ or $\left(\overline{\mathbf{4}},\mathbf{1},\pm\frac{9}{2}\right)$ 
        &
        $\mathbf{443520'}$, and c.c. \\
        \hline
    \end{tabular}
\end{table*}

Unfortunately, if we consider a higher dimensional representation Higgs to form the generalized Z-string as above,  the simple unification of matters in the usual $SO(10)$ GUT becomes impossible because the definition of $U(1)_Y$ becomes different from the
usual one.  
For example, if a VEV of Higgs whose representation is $(\mathbf{1},\mathbf{1},\mathbf{2},9/(2\sqrt{6}))$ under $SU(3)_C\times SU(2)_L\times SU(2)_R\times U(1)_X$, which is included in $\mathbf{1200}$ for example, breaks $SU(2)_R\times U(1)_X$ into $U(1)_Y$,  the spinor  ${\mathbf 16}$ is divided as
\begin{align}
    \label{fermion_decompose_1200}
    \mathbf{16} = &\left(\mathbf{3}, \mathbf{2}, \frac{1}{6}\right) \oplus \left(\mathbf{1}, \mathbf{2}, -\frac{1}{2}\right) \oplus \left(\overline{\mathbf{3}}, \mathbf{1}, -\frac{5}{3}\right) \nonumber\\
    &\oplus \left(\overline{\mathbf{3}}, \mathbf{1}, \frac{4}{3}\right) \oplus \left(\mathbf{1}, \mathbf{1}, 2\right) \oplus \left(\mathbf{1}, \mathbf{1}, -1\right) ,
\end{align}
where we normalize the $U(1)_Y$ charge so that the hypercharge of the doublet quark
becomes 1/6. Since this charge assignment in Eq.(\ref{fermion_decompose_1200}) is different from the unification of matters in the usual $SO(10)$ GUT as
\begin{align}
    \label{fermion_decompose_16}
    \mathbf{16} = &\left(\mathbf{3}, \mathbf{2}, \frac{1}{6}\right) \oplus \left(\mathbf{1}, \mathbf{2}, -\frac{1}{2}\right) \oplus \left(\overline{\mathbf{3}}, \mathbf{1}, -\frac{2}{3}\right) \nonumber\\
    &\oplus \left(\overline{\mathbf{3}}, \mathbf{1}, \frac{1}{3}\right) \oplus \left(\mathbf{1}, \mathbf{1}, 1\right) \oplus \left(\mathbf{1}, \mathbf{1}, 0\right) ,
\end{align}
other representation fields are needed to include the SM quark and leptons in this scenario. It is interesting to build concrete $SO(10)$ GUT models which include the SM
quarks and leptons with stable generalized Z-string, but we think that it is beyond the scope of this paper.

Finally, we should note that if the origins of $SU(N)$ and $U(1)_X$ are independent of each other, we do not have the difficulty which appears  in the above unified models to satisfy the condition (\ref{constraint_couplings}).

%%%%%%%%%%%%%%%%%%%%%%%%%%%%%%%%
\section{discussion and summary}
%%%%%%%%%%%%%%%%%%%%%%%%%%%%%%%%

The embedded strings, which can be cosmic strings, are produced even when the first homotopy group of the moduli space is trivial unlike the topological string such as the N-O string. The classical stability of the embedded string is not guaranteed by the topological features of the moduli space, and it has been studied for the Z-string in the breaking $SU(2)_L\times U(1)_Y\rightarrow U(1)_Q$ by analytically and numerically since 1990s\cite{Vachaspati:1992fi,James:1992zp,La:1993je,Earnshaw:1993yu,Perivolaropoulos:1993gg,Holman:1992rv}, but few for other symmetry breaking\cite{Vachaspati:1992pi}.

We have considered the embedded string in the breaking $SU(N)\times U(1)_X\rightarrow SU(N-1)\times U(1)_Q$ caused by $SU(N)$ fundamental Higgs with $U(1)_X$ charge, which we call the generalized Z-string in this paper. 
We have examined the classical stability of the generalized Z-string by perturbation from the Hamiltonian at tree level. We have found that its stability is determined only by two mass ratios, $m_\phi/m_{\tilde Z}$ and $m_G/m_{\tilde Z}$. Note that it does not depend on $N$ explicitly.   
 This means that 
the region in the parameter space of the two mass ratios, where the generalized Z-string becomes stable, 
can be obtained by that for the Z-string in Ref.\cite{James:1992zp}, which is also given in Fig. \ref{gZstring_stableregion} in this paper.

We have also considered the generalized Z-string in the SUSY $SU(N)\times U(1)_X$ Higgs model and have shown that  
the discussion of the stability of the SUSY generalized Z-strings is exactly the same as in the case of non-SUSY, 
although SUSY models include two Higgs fields. Therefore, the stability is determined by the two mass ratios and  the region in the parameter space of the two mass ratios, in which 
the SUSY generalized Z-string is classically stable, is equivalent to that in the non-SUSY case. It is an extension of the result that we pointed out in Ref.\cite{Kanda:2022xrz}. 

We have applied the stability condition for the generalized Z-string into several models. 
If the origins of $SU(N)$ and $U(1)_X$ are independent of each other, the
stable generalized Z-string just requires much larger mass of the neutral gauge boson than the
charged gauge boson mass. 
Furthermore, we have considered the case in which $SU(N)$ and $U(1)_X$ had been unified into a simple gauge group $G$, that fixes the normalization of the $U(1)_X$.  We have shown that the $U(1)_X$ charge of the Higgs which breaks $SU(N)\times U(1)_X$ into $SU(N-1)\times U(1)_Q$ must be large to obtain the stable generalized Z-string. This requires that higher representation field of $G$ includes the Higgs field. 
We have applied the condition  to several patterns of symmetry breaking and have discussed 
how large representation  Higgs under $G$ is 
needed for formation of the generalized Z-string. 

Our results on the embedded string are based on the potential in tree level. If we take account of the  effective potential, our results must change. For example, the stability
depends not only on the two mass ratios but also on $N$ and/or Yukawa couplings if any.

Our ultimate goal is to test the models beyond the SM through the embedded string.
As the first step, we have clarified what kind of model the embedded string is formed in,
when the $SU(N)\times U(1)_X$ is broken into $SU(N-1)\times U(1)_Q$. If we can know how to observe the cosmic embedded string, the goal will be achieved. For example, if the embedded string has sufficiently long lifetime, the gravitational waves from the cosmic embedded string are expected to be similar to those from the topological string. Thus,
the NANOGrav's result in 2020 can be interpreted as the gravitational waves not only from
the topological string but also from the embedded string. If the embedded string easily decays, the cosmological observables of
the embedded string must be different from those of the topological string. It must be interesting to clarify these issues, but these are beyond the scope of this paper.

\begin{acknowledgments}
    The authors thank Yu Hamada  for useful discussions about numerical calculation.  
    This work is supported in part by the Grant-in-Aid for Scientific Research 
      from the Ministry of Education, Culture, Sports, 
     Science and Technology in Japan  No.~19K03823(N.M.).
\end{acknowledgments}

\appendix

\section{}

In this appendix, we show that only one perturbation mode determines
the stability of the generalized Z-string. This argument is essentially the same as discussed in Ref.\cite{James:1992zp}.

We have shown that only perturbations of charged fields may destabilize the generalized Z-string in the $SU(N)\times U(1)_X$ Higgs model. The variation of the energy linear density is given as
\begin{widetext}
\begin{align}
    \label{mu_k}
    \mu_k = 
    &\int d^2x \left[ 2\left\{ \left(\nabla\times\vec{G}^{k+}\right)_z -\frac{ig_N^2}{\alpha_N}\frac{z}{r}G^{k+}_r \right\} \left\{ \left(\nabla\times\vec{G}^{k-}\right)_z + \frac{ig_N^2}{\alpha_N}\frac{z}{r}G^{k-}_r \right\} + \frac{2ig_N^2}{\alpha_N} \frac{z'}{r} \left(\vec{G}^{k+}\times\vec{G}^{k-}\right)_z \right.\nonumber\\
    &\hspace{13mm} + \left| \partial_r \phi_{c,k} \right|^2 + \frac{1}{r^2}\left| \left(\partial_\theta + i\frac{\alpha_N}{2} \left(\frac{\cos^2\theta_G}{N-1} - \sin^2\theta_G\right) z\right) \phi_{c,k} \right|^2 \nonumber\\
    &\hspace{13mm} -ig_N \left\{ \left(f' G^{k-}_r -if \frac{1-\frac{\alpha_N}{2}z}{r} G^{k-}_\theta\right)e^{-i\theta} \phi_{c,k} \right.\nonumber\\
    &\hspace{23mm} \left. + f e^{i\theta} \left(G^{k+}_r \partial_r\phi_{c,k}^* + \frac{G^{k+}_\theta}{r}\left(\partial_\theta - i\frac{\alpha_N}{2}\left(\frac{\cos^2\theta_G}{N-1} - \sin^2\theta_G\right)z\right)\phi_{c,k}^*\right) \right\} \nonumber\\
    &\hspace{13mm} +ig_N \left\{ \left(f' G^{k+}_r +if \frac{1-\frac{\alpha_N}{2}z}{r} G^{k+}_\theta\right)e^{i\theta} \left( \phi_{c,k}\right)^* \right.\nonumber\\
    &\hspace{23mm} \left. + f e^{-i\theta} \left(G^{k-}_r \partial_r\phi_{c,k} + \frac{G^{k-}_\theta}{r}\left(\partial_\theta + i\frac{\alpha_N}{2}\left(\frac{\cos^2\theta_G}{N-1} - \sin^2\theta_G\right)z\right)\phi_{c,k}\right)  \right\} \nonumber\\
    &\hspace{13mm} \left.+ g_N^2f^2 G^{k+}_{\bar{i}}G^{k-}_{\bar{i}} + 2\lambda\left(f^2 -v^2\right)|\phi_{c,k}|^2 \right] ,
\end{align}
\end{widetext}
where $k$ is an any integer from 1 to $N-1$.

First, we consider the CP-invariance of $\mu_k$. 
Thus, $\mu_k$ can be divided as $\mu_k = \mu_+ + \mu_-$, where $\mu_+$ is composed of CP-even perturbations and $\mu_-$ is composed of CP-odd ones. 
In the expansion of $\phi_{c,k}, \vec{G}^{k+}$ and $\vec{G}^{k-}$ as 
\begin{widetext}
\begin{align}
    \label{charged_mode_exsp1}
    &\phi_{c,k}(x) = \sum_{m=-\infty}^\infty \phi_m(r) (i)^m e^{im\theta} \\
    \label{charged_mode_exsp2}
    &\vec{G}^{k+}(x) = \sum_{m=-\infty}^\infty \left[ G_m(r) (i)^me^{im\theta}\vec{e}_r + \frac{\xi_m(r)}{r} (i)^{m+1}e^{im\theta}\vec{e}_\theta \right] \\
    \label{charged_mode_exsp3}
    &\vec{G}^{k-}(x) = \sum_{m=-\infty}^\infty \left[ G^*_{-m}(r) (i)^me^{im\theta}\vec{e}_r - \frac{\xi^*_{-m}(r)}{r} (i)^{m+1}e^{im\theta}\vec{e}_\theta \right], 
\end{align}
\end{widetext}
where $\vec{G}^{k-}=\left(\vec{G}^{k+}\right)^*$,
CP-even modes and CP-odd modes are given as
\begin{align}
    CP\mbox{-even modes:}\quad &\mbox{Re}\left[\phi_m(r)\right], \mbox{Re}\left[G_m(r)\right], \mbox{Re}\left[\xi_m(r)\right] , \\
    CP\mbox{-odd modes:}\quad &\mbox{Im}\left[\phi_m(r)\right], \mbox{Im}\left[G_m(r)\right], \mbox{Im}\left[\xi_m(r)\right]
\end{align}
because  CP transformation for fields on $(r, \theta)$ plane is equivalent to taking the complex conjugate of them and coordinate transformation such that $\theta\rightarrow\pi-\theta$.

On the other hand, since an $U(1)_Q$ gauge transformation, which just multiplies $\pm i$ to the fields  $\phi_{c,k}, \vec{G}^{k+}$ and $\vec{G}^{k-}$, exchanges the CP-even modes and CP-odd modes, 
we can conclude that the instabilities made by $\mu_+$ and $\mu_-$ are equivalent and it is sufficient to examine only one of them. Hereinafter, we consider about only $\mu_+$, in other words, we assume that $\phi_m(r), G_m(r)$ and $\xi_m(r)$ in Eq.(\ref{charged_mode_exsp1}, \ref{charged_mode_exsp2}, \ref{charged_mode_exsp3}) are real.

Substituting the expanded form of $\phi_{c,k}, \vec{G}^{k+}$ and $\vec{G}^{k-}$ into $\mu_+$ and integrating by $\theta$, we obtain that
\begin{widetext}
\begin{align}
    \label{mu_c1_n=1}
    \mu_+
    &=2\pi \int rdr\sum_m
    \left[ \frac{2}{r^2} \left({\xi_{m-1}}' - \left(m-\left(1-\frac{g_N^2}{\alpha_N}z\right)\right)G_{m-1}\right)^2
    + \frac{4g_N^2}{\alpha_N}\frac{z'}{r^2}G_{m-1}\xi_{m-1}
    \right. \nonumber\\
    &\hspace{30mm} + ({\phi_m}')^2 + \frac{1}{r^2} \left(m+ \frac{\alpha_N}{2}\left(\frac{1}{N-1}\cos^2\theta_G - \sin^2\theta_G \right)z\right)^2\phi_m^2 \nonumber\\
    &\hspace{30mm} +2g_N \left\{ f'G_{m-1}\phi_m - f\frac{1-\frac{\alpha_N}{2}z}{r^2}\xi_{m-1}\phi_m - fG_{m-1}{\phi_m}' \right.\nonumber\\
    &\hspace{45mm} \left. - \frac{f}{r^2} \left(m+ \frac{\alpha_N}{2}\left(\frac{1}{N-1}\cos^2\theta_G - \sin^2\theta_G \right)z\right)\xi_{m-1}\phi_m\right\} \nonumber\\
    &\hspace{30mm} \left.+ g_N^2f^2 \left\{G_{m-1}^2 + \frac{\xi_{m-1}^2}{r^2}\right\} + 2\lambda\left(f^2 - v^2\right) \phi_m^2 \right] .
\end{align}
\end{widetext}
The fourth term of the integrand in Eq.(\ref{mu_c1_n=1}) diverges positively in $r\sim 0$ when $m\neq 0$, thus we set $m=0$ because it is most likely to be negative. If we write the $m=0$ modes simply as 
\begin{align}
    \chi(r) \equiv \phi_0(r),\quad G(r) \equiv G_{-1}(r),\quad \xi(r) \equiv \xi_{-1}(r) ,
\end{align}
the $m=0$ part of $\mu_+$ is given as
\begin{widetext}
\begin{align}
    \label{mu_c1_n0_m0}
    &\mu_0 = 2\pi \int rdr
    \left[ \frac{2}{r^2} \left({\xi}' + \left(1-\frac{g_N^2}{\alpha_N}z\right)G\right)^2
    + \frac{4g_N^2}{\alpha_N}\frac{z'}{r^2}G\xi + (\chi')^2 + \frac{\alpha_N^2}{4r^2} \left(\frac{1}{N-1}\cos^2\theta_G - \sin^2\theta_G \right)^2z^2 \chi^2 \right.\nonumber\\
    &\hspace{30mm} +2g_N \left\{ \left(f'\chi-f\chi'\right)G - \frac{f}{r^2} \left(1+\frac{\alpha_N}{2}\left(\frac{1}{N-1}\cos^2\theta_G - \sin^2\theta_G -1\right)\right) \chi\xi \right\} \nonumber\\
    &\hspace{30mm} \left.+ g_N^2f^2 \left\{G^2 + \frac{\xi^2}{r^2}\right\} + 2\lambda\left(f^2 - v^2\right) \chi^2 \right] .
\end{align}
\end{widetext}

Since there are no terms having derivative of $G(r)$ in Eq.(\ref{mu_c1_n0_m0}), we can transform all terms depending on $G(r)$ in the integrand of $\mu_0$ into a complete square as
\begin{widetext}
\begin{align}
    &\left(g_N^2 + \frac{2}{r^2} \left(1-\frac{g_N^2}{\alpha_N}z\right)^2 \right) G^2
    + \left\{ \frac{4}{r^2}\left(1-\frac{g_N^2}{\alpha_N}z\right)\xi' + \frac{4g_N^2}{\alpha_N}\frac{z'}{r^2} \xi + 2g_N\left(f'\chi - f\chi'\right) \right\} G \nonumber\\
    &= \frac{2}{r^2}P_N \left[G + \frac{1}{P_N} \left\{ \left(1-\frac{g_N^2}{\alpha_N}z\right)\xi' + \frac{g_N^2}{\alpha_N}z' \xi + \frac{g_Nr^2}{2}\left(f'\chi - f\chi'\right) \right\} \right]^2 \nonumber\\
    & - \frac{2}{r^2P_N} \left\{ \left(1-\frac{g_N^2}{\alpha_N}z\right)\xi' + \frac{g_N^2}{\alpha_N}z' \xi + \frac{g_Nr^2}{2}\left(f'\chi - f\chi'\right) \right\}^2 ,
\end{align}
\end{widetext}
where 
\begin{align}
    P_N \equiv \frac{g_N^2r^2f^2}{2} + \left(1- \frac{g_N^2}{\alpha_N} z\right)^2 .
\end{align}
This implies that the perturbation $G(r)$ does not makes a negative variation of the energy linear density. Therefore, we assume that $G(r)$ satisfies 
\begin{align}
    \label{constraint_G}
    G = - \frac{1}{P_N} \left\{ \left(1-\frac{g_N^2}{\alpha_N}z\right)\xi' + \frac{g_N^2}{\alpha_N}z' \xi + \frac{g_Nr^2}{2}\left(f'\chi - f\chi'\right) \right\} ,
\end{align}
and ignore it.

To summarize the calculations so far, the variation of the energy linear density we consider here is given as
\begin{widetext}
\begin{align}
    \mu_0 &= \int rdr \left[
    \frac{2}{r^2} ({\xi}')^2 + ({\chi}')^2 + \frac{\alpha_N^2}{4r^2} \left(\frac{1}{N-1}\cos^2\theta_G - \sin^2\theta_G\right)^2 z^2 \chi^2 \right. \nonumber\\
    &\hspace{15mm} -2g_N \frac{f}{r^2} \left( 1 + \frac{\alpha_N}{2} \left( \frac{1}{N-1}\cos^2\theta_G - \sin^2\theta_G -1 \right) \right) \xi \chi
    +\frac{g_N^2f^2}{r^2} \xi^2 + 2\lambda(f^2-v^2)\chi^2 \nonumber\\
    &\hspace{15mm} \left. - \frac{2}{r^2P} \left\{ \left(1-\frac{g_N^2}{\alpha_N}z\right)\xi' + \frac{g_N^2}{\alpha_N}z' \xi + \frac{g_Nr^2}{2}\left(f'\chi - f\chi'\right) \right\}^2 \right] ,
\end{align}
\end{widetext}
where we have only two perturbation modes $\chi(r)$ and $\xi(r)$. Moreover, 
since one linear combination of them corresponds to a gauge transformation,  the physical degree of freedom of perturbations becomes only one as follows.
By considering an infinitesimal $SU(N)$ gauge transformation which is calculated as
\begin{align}
    &\phi(x) \rightarrow \phi(x) + ig_N\Lambda (x) \phi(x) , \nonumber\\
    &G^a_\mu(x) \rightarrow G^a_\mu(x) + D_\mu \Lambda^a(x) ,
\end{align}
where $\Lambda(x) \equiv \Lambda^a(x)T^a$ is a real $\mathfrak{su}(N)$ valued function and 
\begin{align}
    D_\mu \Lambda^a(x) \equiv \partial_\mu \Lambda^a(x) -ig_N \left[G_\mu(x), \Lambda(x)\right]^a .
\end{align}
We set $\Lambda(x)$ as
\begin{align}
    \Lambda(x) = &s(r) \sin\theta T^{(N-1)^2+2(k-1)} \nonumber\\
    &- s(r) \cos\theta T^{(N-1)^2+2k-1} ,
\end{align}
where $s(r)$ is a smooth function, and consider the infinitesimal $SU(N)$ gauge transformation of the generalized Z-string solution in Eq.(\ref{gZstring_ansatz}). As a result of calculations, we find that it is equivalent to the following perturbations\footnote{Note $G(r)$ in Eq.(\ref{gaugetrf_perturbation}) satisfies the condition in Eq.(\ref{constraint_G}).},
\begin{align}
    \label{gaugetrf_perturbation}
    &\chi(r) = -g_Nf(r)s(r), \quad \xi(r) = \left(1-\frac{g_N^2}{\alpha_N}z(r)\right)s(r), \nonumber\\
    &G(r) = -s'(r) .
\end{align}
Because they are unphysical perturbations, they do not change the form of $\mu_0$. Hence perturbations that are proportional to Eq.(\ref{gaugetrf_perturbation}) must vanish in the integrand of $\mu_0$ and only the perturbations perpendicular to them which are denoted as
\begin{equation}
    \zeta_N(r) \equiv \left(1-\frac{g_N^2}{\alpha_N}z\right) \chi(r) + g_Nf \xi(r)
\end{equation}
are remained. After some calculations, we obtain that
\begin{widetext}
\begin{align}
    \label{app_delta_mu}
    \mu_0 &= \int rdr \left[ \frac{1}{P_N} ({\zeta_N}')^2 + \left\{ \frac{2S_N}{r^2}f^2 + \frac{1}{r}\frac{d}{dr}\left(\frac{rf'}{P_Nf}\right) + \frac{(f')^2}{P_Nf^2} \right\}\zeta_N^2 \right] \nonumber\\
    &= \int rdr \, \zeta_N \left[ -\frac{1}{r} \frac{d}{dr}\left(\frac{r}{P_N}\frac{d}{dr}\right) + \left\{ \frac{2S_N}{r^2f^2} + \frac{1}{r}\frac{d}{dr}\left(\frac{rf'}{P_Nf}\right) + \frac{(f')^2}{P_Nf^2} \right\} \right]\zeta_N \nonumber\\
    &\equiv \int rdr \, \zeta_N \mathcal{O}_N \zeta_N ,
\end{align}
\end{widetext}
where
\begin{align}
    S_N \equiv r\frac{d}{dr}\left(\frac{1}{rP_N}\left(1-\frac{g_N^2}{\alpha_N}z\right)\frac{z'}{\alpha_N}\right) + \frac{f^2}{2} -\frac{1}{P_N}\frac{g_N^2}{\alpha_N^2}(z')^2 .
\end{align}
Hence we can find that only the perturbation mode $\zeta_N(r)$ may be able to destabilize the generalized Z-string. 

If $\mathcal{O}_N$ has a negative eigenvalue, there must be a perturbation which makes $\mu_0$ negative. Hence the parameter region such that $\mathcal{O}_N$ does not have a negative eigenvalue is the region that the generalized Z-string solutions are stable. If we take the nondimensionalization in (\ref{normalize_SUN}), $\mathcal{O}_N$ becomes
\begin{align}
    \label{O_gZstring}
    \mathcal{O}_N &= \frac{m_{\tilde{Z}}^2}{4} \left[ -\frac{1}{R} \frac{d}{dR}\left(\frac{R}{\tilde{P}_N}\frac{d}{dR}\right) \right.\nonumber\\
    &\hspace{20mm}\left.+ \left\{ \frac{2\tilde{S}_N}{R^2F^2} + \frac{1}{R}\frac{d}{dR}\left(\frac{RF'}{\tilde{P}_NF}\right) + \frac{(F')^2}{\tilde{P}_NF^2} \right\} \right] \nonumber\\
    &\equiv \frac{m_{\tilde{Z}}^2}{4} \tilde{\mathcal{O}}_N ,
\end{align}
where
\begin{align}
    &\tilde{P}_N \equiv P_N = 2 \frac{m_G^2}{m_{\tilde{Z}}^2} R^2 F^2 + \left(1 - 2\frac{m_G^2}{m_{\tilde{Z}}^2} Z\right)^2 \\
    &\tilde{S}_N \equiv \frac{S_N}{v^2} = \frac{R}{2}\frac{d}{dR} \left( \frac{Z'}{R\tilde{P}_N}\left(1-2\frac{m_G^2}{m_{\tilde{Z}}^2} Z\right) \right) \nonumber\\
    &\hspace{20mm}+ \frac{F^2}{2} - \frac{1}{\tilde{P}_N}\frac{m_G^2}{m_{\tilde{Z}}^2} (Z')^2 .
\end{align}
We can see that eigenvalues of $\tilde{\mathcal{O}}_N$ depends on $F(R)$, $Z(R)$ and $m_G/m_{\tilde{Z}}$. Since  the shape of $F(R)$ and $Z(R)$ are determined by $m_\phi/m_{\tilde{Z}}$, we can conclude that the stability of the generalized Z-string depends on $m_\phi/m_{\tilde{Z}}$ and $m_G/m_{\tilde{Z}}$ as in section 3.

%%%%%%%%%%%%%%%%%%%%%%%%%%%%%

\end{document}